%Paper: hep-th/9306052
%From: JXLU@crnvma.cern.ch
%Date: Thu, 10 Jun 93 22:00:57 SET

%
%input harmac
%

\input harvmac.tex

%%%%%%%%%%%%%%%%%%%%%%%%%%REFERENCES FOR BLACK AND SUPER%%%%%%%%%%%%%%%%%%%%%%
\lref\dufsp{M. J. Duff, E. Sezgin and C. N. Pope, ``Supermembranes and Physics
in $2 + 1$ Dimensions'' (World Scientific, 1990).}

\lref\duflbs{M. J. Duff and J. X. Lu, Nucl. Phys. {\bf B390} (1993) 276.}

\lref\hors{G. T. Horowitz and A. Strominger, Nucl. Phys. {\bf B360} (1991)
197.}

\lref\gids{S. Giddings and A. Strominger, Phys. Rev. Lett. {\bf 67}
(1991) 2930.}

\lref\hor{G. T. Horowitz, ``The dark side of string theory: black holes
and black strings", UCSBTH-92-32.}

\lref\dufkl{M. J. Duff, R. Khuri and J. X. Lu, Nucl. Phys.
{\bf B377} (1992) 281.}

\lref\guv{R. Guven, ``Black $p$-brane solutions of $D = 11$
supergravity",  Bogazici University preprint (1991).}

\lref\grel{R. Gregory and R. Laflamme, ``Black strings and $p$-branes
are unstable", EFI-93-02.}

\lref\huglp{J. Hughes, J. Liu and J. Polchinski, Phys. Lett. {\bf B180}
(1986) 370.}

\lref\tow{P. K. Townsend, Phys. Lett. {\bf B202} (1988) 53.}

\lref\str{A. Strominger, Nucl. Phys. {\bf B343} (1990) 167.}

\lref\dufldu{M. J. Duff and J. X. Lu, Nucl. Phys. {\bf B354} (1991) 129.}

\lref\duflfb{M. J. Duff and J. X. Lu, Nucl. Phys. {\bf B354} (1991) 141.}

\lref\dufllo{M. J. Duff and J. X. Lu, Nucl. Phys. {\bf B357} (1991) 534.}

\lref\hars{J. Harvey and A. Strominger, Phys. Rev. Lett. {\bf 66} (1991) 549.}

\lref\duflss{M. J. Duff and J. X. Lu, Phys. Rev. Lett. {\bf 66} (1991) 1402.}

\lref\calhsone{C. Callan, J. Harvey and A. Strominger, Nucl. Phys. {\bf B359}
(1991) 611.}

\lref\calhstwo{C. Callan, J. Harvey and A. Strominger, Nucl. Phys. {\bf B367}
(1991) 60.}

\lref\calhsthree{C. Callan, J. Harvey and A. Strominger, ``Supersymmetric
string
solitons'', EFI-91-66, hep-th/9112030.}

\lref\duflsfdu{M. J. Duff and J. X. Lu, Class. and Quantum. Grav. {\bf 9}
(1991) 1.}

\lref\dufltb{M. J. Duff and J. X. Lu, Phys. Lett. {\bf B273} (1991) 409.}

\lref\senone{A. Sen, ``Black holes and solitons in string theory'',
TIFR-TH-92-57, hep-th/9210050.}

\lref\sentwo{A. Sen, ``Macroscopic charged heterotic string'', TIFR-TH-92-29,
 hep-th/9206016.}

\lref\senthree{A. Sen, ``Electric magnetic duality in string theory'',
 TIFR-TH-92-41, hep-th/9207053.}

\lref\duf{M. J. Duff, Class. and Quantum Grav. {\bf 5} (1988) 189.}

\lref\schs{ J. H. Schwarz and A. Sen, "Duality symmetric actions",
NSF-ITP-93-46, CALT-68-1863, TIFR-TH-93-19.}

\lref\sch{J. H. Schwarz, Nucl. Phys. {\bf B226} (1983) 269.}

\lref\nep{R. Nepomechie, Phys. Rev. {\bf D31} (1984) 1921.}

\lref\tei{C. Teitelboim, Phys. Lett. {\bf B167} (1986) 69.}

\lref\gibp{G. W. Gibbons and M. J. Perry, Nucl. Phys. {\bf B248} (1984)
629.}

\lref\gibw{G. W. Gibbons and D. L. Wiltshire, Ann. Phys. {\bf 167} (1986)
201.}

\lref\rey{S. J. Rey, Phys. Rev. {\bf D 43} (1991) 526.}

\lref\gibm{G. W. Gibbons and K. Maeda, Nucl. Phys. {\bf B298} (1988) 741.}

\lref\calk{C. Callan and R. Khuri, Phys. Lett. {\bf B201} (1991) 363.}

\lref\gresw{M. Green, J. Schwarz and E. Witten, Superstring Theory
(C.U.P. 1987).}

\lref\berddv{E. Bergshoeff, M. de Roo, B. de Wit and P. van Nieuwenhuizen,
Nucl. Phys. {\bf B195} (1982) 97.}

\lref\cham{G. F. Chapline and N. Manton, Phys. Lett. {\bf B120} (1983) 105.}

\lref\cha{A. H. Chamseddine, Nucl. Phys. {\bf B185} (1981) 403.}

\lref\nictv{H. Nicolai, P. K. Townsend and P. van Nieuwenhuizen, Lett.
Nuovo Cimento {\bf 30} (1981) 315.}

\lref\rom{L. Romans, Nucl. Phys. {\bf B276} (1986) 71.}

\lref\shatw{A. Shapere, S. Trivedi and F. Wilczek, Mod. Phys. Lett. {\bf A6}
(1991) 2677.}

\lref\kallopv{R. Kallosh, A. Linde, T. Ortin, A. Peet and A. van Proeyen,
Phys. Rev. {\bf D46} (1992) 5278.}

\lref\lu{J. X. Lu, "ADM masses for black strings and $p$-branes",
 CERN-TH.6877/93, hep-th/9304159.}

\lref\berst {E. Bergshoeff, E. Sezgin and P. Townsend, Phys. Lett. {\bf B189}
 (1987)  75.}

\lref\achetw {A. Achucarro, J. Evans, P. Townsend and D. Wiltshire, Phys.
Lett. {\bf B198}  (1987)  441.}

\lref\dabghr {A. Dabholkar, G. W. Gibbons, J. A. Harvey and F. Ruiz Ruiz,
Nucl. Phys. {\bf B340}  (1990)  33.}

\lref\dufhis {M. J. Duff, P. S. Howe, T. Inami and K. Stelle, Phys. Lett.
{\bf B191}  (1987)  70.}

\lref\dufs {M. J. Duff and K. Stelle, Phys. Lett. {\bf B253}  (1991)  113.}

\lref\han {S. K. Han, J. K. Kim, I. G. Koh, and Y. Tanii, Phys. Rev. {\bf D34}
(1986) 553.}

\lref\sala {A. Salam and E. Sezgin, ``Supergravities in Diverse Dimensions'',
(North
Holland/World Scientific 1989).}

\def\dg{\hbox{$^\dagger$}}
\def\ddg{\hbox{$^\ddagger$}}

\def\half{{1\over2}}

%For more complicated situations, substitute for {\it either\/} argument:
%\Title{\vbox{\baselineskip12pt\hbox{HUT$p$-88/A000}\hbox{SLAC-PUB 88-001}
%               \hbox{photocopy at own risk}}}
%{\vbox{\centerline{This title is too long to fit}
%       \vskip2pt\centerline{comfortably on one line*}}}
%   \footnote{}{*optional footnote on title}

\Title{\vbox{\baselineskip12pt \hbox{CERN-TH.6675/93}\hbox{CTP/TAMU-54/92}}}
{\vbox{\centerline{\bf BLACK AND SUPER P-BRANES IN DIVERSE DIMENSIONS}}}

\centerline {M. J. Duff{\footnote\dg{Work supported in part by NSF grant
PHY-9106593}}}
\centerline{\it{Center for Theoretical Physics, Physics Department,
Texas A \& M University}}
\centerline{\it{College Station, TX 77843}}
\medskip
\centerline{J. X. Lu {\footnote\ddg{Supported in part by
a World Laboratory Fellowship}}}
\centerline{\it{CERN, Theory Division, CH-1211, Geneva 23, Switzerland}}
\bigskip

\baselineskip = 10pt
\centerline{\bf Abstract}

We present a generic Lagrangian, in arbitrary spacetime dimension $D$,
describing the interaction of a dilaton, a graviton and an antisymmetric
tensor
of arbitrary rank $d$.  For each $D$~and~$d$, we find ``solitonic'' black
$\tilde{p}$-brane solutions where $\tilde{p} = \tilde{d} - 1$~and~$\tilde d
= D - d - 2$.  These
solutions display a spacetime singularity surrounded by an event horizon, and
are
characterized by a mass per unit $\tilde p$-volume, ${\cal M}_{\tilde{d}}$, and
topological ``magnetic'' charge $g_{\tilde{d}}$, obeying
$\kappa {\cal M}_{\tilde{d}} \geq g_{\tilde{d}}/\sqrt{2}$.  In the extreme
limit
$\kappa {\cal M}_{\tilde{d}} = g_{\tilde{d}}/\sqrt{2}$, the singularity and
event horizon
coalesce.  For specific values of $D$~and~$d$, these extreme solutions also
exhibit supersymmetry and may be identified with previously classified
heterotic, Type IIA and Type IIB super $\tilde p$-branes.  The theory also
admits
elementary $p$-brane solutions with ``electric'' Noether charge $e_d$, obeying
the Dirac quantization rule $e_d g_{\tilde{d}} = 2\pi n$, $n =$~integer.  We
also present the Lagrangian describing the theory dual to the original theory,
whose antisymmetric tensor has rank $\tilde{d}$ and for which the roles of
topological and elementary solutions are interchanged.  The super $p$-branes
and their duals are mutually non-singular.  As special cases of our general
solution we recover the black $p$-branes of Horowitz and Strominger $(D = 10)$,
Guven $(D = 11)$ and Gibbons et al $(D = 4)$, the $N = 1$, $N = 2a$~and~$N =
2b$ super-$p$-branes of Dabholkar et al $(4 \leq D \leq 10)$, Duff and
Stelle $(D = 11)$, Duff and Lu $(D = 10)$ and Callan, Harvey and Strominger $(D
= 10)$, and the axionic instanton of Rey $(D = 4)$.  In particular, the
electric/magnetic duality of Gibbons and Perry in $D = 4$ is seen to be a
consequence of particle/sixbrane duality in $D = 10$.  Among the new solutions
is a self-dual superstring in $D = 6$.
\bigskip

\noindent
CERN-TH.6675/93

\Date{June 1993}
\vfill\eject
\baselineskip=18pt

\newsec{\bf Introduction}
Supersymmetric extended objects \dufsp\ are
interesting for a variety of reasons.  First, they correspond to the extreme
mass $=$ charge limit \duflbs\ of the black $p$-branes
\refs{\hors,\gids,\hor,\guv,\grel}, which are higher
dimensional analogues of
black holes.  These super $p$-branes, stable by virtue of the supersymmetry
which emerges in
this limit, might thus describe the end point of Hawking radiation.  Secondly,
they emerge
as topological defects \refs{\huglp,\tow} of supersymmetric field theories,
and might thus have
interesting cosmological consequences.  In particular they provide soliton
solutions of $N= 1$ supergravity-Yang-Mills, $N = 2A$ supergravity and
$N = 2B$ supergravity in $D = 10$
\refs{\dabghr,\str,\dufs,\dufldu,\duflfb,\dufllo,\hars,\duflss,\calhsone,
\calhstwo,\calhsthree,\duflsfdu,\dufltb,\senone,\sentwo,\senthree} which
are the field theory limits of the $D = 10$ heterotic, Type IIA and  Type IIB
superstrings, respectively.
 In some cases, one can show that they correspond,
in fact,
to exact conformal field theories \refs{\calhsone,\calhstwo,\calhsthree,\gids}
 and
must therefore be taken just as
seriously by
string theorists as magnetic monopoles are by grand unified theorists:  one
cannot buy
superstrings without buying super $p$-branes in the same package!  The final,
more
speculative, but perhaps most intriguing reason, is the possibility that they
 may provide a
dual description of superstrings.  For example, there is a mounting body of
evidence to
suggest that in $D = 10$, the heterotic superstring is dual to the heterotic
fivebrane \refs{\duf,\str} with the strongly coupled string corresponding to
the weakly coupled
fivebrane \refs{\str,\dufldu}.  The study of super $p$-branes might thus throw
light on the strong coupling regime of string theory.

The plan of the paper is as follows:  In section 2, we write down a general
action in $D$
spacetime dimensions describing the interaction of an antisymmetric tensor
potential of
rank $d$ with gravity and a dilaton.  We allow these fields to couple to an
elementary
$d$-dimensional extended object, (a ``$p$-brane'', with $d = p + 1$) and
define an
``electric'' Noether charge associated with it.  In section 3, we show how the
combined
field equations admit solutions describing such elementary objects, in much
the
same way
as Dabholkar et al \dabghr\ showed how an elementary string emerges as a
solution of
supergravity coupled to a string $\sigma$-model source.  As described in
section 4, one
may establish a Bogolmol'nyi bound between mass per unit $p$-volume ${\cal
M}_d$ of the  $p$-brane
 and the
Noether charge $e_d$, and demonstrate that these elementary solutions saturate
the  bound, and
are thus seen to be classically stable.  One may also demonstrate a
``no-static-force''
condition by showing that the mutual gravitational-dilaton attraction of two
such $p$-branes
of the same orientation is exactly cancelled by an equal and opposite
contribution from
the antisymmetric tensor.  This permits the construction of stable
multi-$p$-brane solutions.

In addition to the singular elementary $(d - 1)$-brane solutions carrying
non-zero
``electric'' Noether charge $e_d$, the theory also admits non-singular soliton
$(\tilde{d}- 1$)-brane solutions, where $\tilde{d} = D - d - 2$.  As described
in section 5, these
solutions are dual to the elementary solutions and carry a non-zero
``magnetic''
topological charge $g_{\tilde{d}}$, obeying the Dirac quantization rule
\refs{\nep,\tei}
\eqn\diracquan{e_d g_{\tilde{d}} = 2\pi n, \qquad  n = {\rm integer}.}

In section 6 we consider the theory dual to the theory of section 2, for which
the roles
of antisymmetric tensor field equations and Bianchi identities, and hence
electric and
magnetic charges, are interchanged.  This leads to a relation between the loop
expansion
parameter ${\rm g}_d$ of the $(d - 1)$-brane and the loop expansion parameter
${\rm g}_{\tilde{d}}$ of
the $(\tilde{d} - 1)$-brane.  We find
\eqn\looparam{{\rm g}_d^d = 1/{\rm g}_{\tilde{d}}^{\tilde{d}},}
thus confirming that strongly coupled $(d - 1)$ branes correspond to weakly
coupled $(\tilde{d} - 1)$-branes and vice versa.  The question of duality at
higher orders in this loop expansion is considered  in section 7 where we
generalize the $D = 10$ string/fivebrane results of  \dufllo\ to arbitrary
$d$~and~$\tilde{d}$.

Thus far, our discussion has been valid at arbitrary spacetime dimension $D$
and worldvolume dimension $d$.  The important case of $D = 10$ is treated in
section 8 where
recover as special cases of our general solution the $N = 1$ superstring
\dabghr\ and $N = 1$
superfivebrane \duflfb, the Type IIA superparticle \duflbs,
superstring \calhsone, supermembrane \duflbs, superfourbrane \duflbs,
superfivebrane \calhsone\ and supersixbrane \duflbs and the Type IIB
superstring \dabghr, self-dual superthreebrane \dufltb\ and superfivebrane
\calhsone.  Another
special case, the $D = 11$ supermembrane \dufs\ is recovered in section 10,
from which the
$D = 10$ superstring follows by simultaneous dimensional reduction \dufhis\
 of spacetime and
worldvolume.  $D = 6$ is of special interest because in this dimension a
string
is dual
to another string.  This could be either the usual strong/weak duality
$\phi \rightarrow - \phi$ or else, in analogy with the threebrane in $D = 10$,
 via a
\underbar{self-duality}, and in section 11 we present a discussion of the
$D = 6$ self-dual superstring.

In section 12 and 13, we turn to a case of obvious interest:  $D = 4$.
First of all
in section 12, we recall the ``electric'' particle/``magnetic'' monopole
duality of
Gibbons and Perry \gibp\ and show how it follows a consequence of
$0$-brane/$6$-brane
duality
in $D = 10$.  Secondly, in section 13, we recover another special case of our
general
solution the $D = 4$, $\tilde{d} = 0$ ``axionic instanton'' \rey.

Solutions with ${\cal M}_{\tilde{d}} \geq {1\over \sqrt{2}} g_{\tilde{d}}$ are
discussed
in section 14.  These solutions exhibit singularities shielded by an event
horizon.  As
special cases, we recover the $D = 10$ black $p$-branes $(p = 0, \ldots 6)$ of
 Horowitz and
Strominger \hors, the $D = 11$ black $p$-branes $(p = 2, 5)$ of Guven \guv\ and
the $D = 4$ black hole $(p = 1)$ of Gibbons et al \refs{\gibp,\gibw}.  Finally
in
section 15,  we generalize the
results of \dufkl\ to show that although the elementary $(d - 1)$-brane is a
singular solution of $(d -1)$-brane theory, it is a non-singular soliton
solution of $(\tilde{d} - 1)$-brane
theory, and vice-versa.

\newsec{\bf  General equations for arbitrary ($d, D$)}

Consider an antisymmetric tensor potential of rank $d,~A_{M_1 M_2 \ldots M_d}$,
in $D$
spacetime dimensions ($M = 0, 1, \ldots (D-1)$) interacting with gravity,
$g_{MN}$, and
the dilaton, $\phi$, via the action
\eqn\lgeact{I_D (d) = {1\over 2\kappa^2} \int d^D x \sqrt{-g} \Bigg(R -
{1\over 2} (\partial\phi)^2 -
{1\over 2(d+1)!} e^{-\alpha (d) \phi} F^2_{d+1}\Bigg),}
where the rank ($d+1$) field strength $F_{d+1}$ is given by
\eqn\defF{F_{d+1} = dA_d\hfil,}
and $\alpha (d)$ is an, as yet undetermined, constant.  Special cases of
this action
 have
been considered before in the context of classical solutions
\gibm\ [2--25].  Here we
keep
 both
$D$ and $d$ arbitrary.  We allow these fields to couple to an elementary
$d$-dimensional
extended object (a ``($d-1$)-brane'') whose trajectory is given by
$X^M (\xi^i)~(i = 0, 1,\ldots (d-1))$, worldvolume metric by
$\gamma_{ij} (\xi)$, and tension by $T_d$,
 via the
action
\eqn\gepbact{\eqalign{S_d = T_d \int d^d\xi \Bigg(&-{1\over 2} \sqrt{-\gamma}
 \gamma^{ij}
\partial_i X^M \partial_j X^N g_{MN} e^{\alpha(d)\phi/d} + {(d-2)\over 2}
\sqrt{-\gamma}\cr
&-{1\over d!} \varepsilon^{i_1 i_2 \ldots i_d} \partial_{i_1} X^{M_1}
\partial_{i_2}
X^{M_2} \ldots \partial_{i_d} X^{M_d} A_{M_1 M_2 \ldots M_d}\Bigg).\cr}}
The $\phi$ dependence is chosen so that under the rescaling
\eqn\resclaw{\eqalign{g_{MN}&\rightarrow \lambda^{2d/(D-2)} g_{MN},\cr
A_{M_1 M_2 \ldots M_d}&\rightarrow \lambda^d A_{M_1 M_2 \ldots M_d},\cr
e^{\phi}&\rightarrow \lambda^{2d (D-d-2)/(D-2)\alpha (d)} e^{\phi},\cr
\gamma_{ij}&\rightarrow \lambda^2 \gamma_{ij},\cr}}
both actions scale the same way
\eqn\actscal{\eqalign{I_D (d)&\rightarrow \lambda^d I_D (d),\cr
S_d&\rightarrow \lambda^d S.\cr}}
The field equations and Bianchi identities of the $A$ field may be written
\eqn\fieldeq{d^{\ast} (e^{-\alpha (d)\phi} F) = 2\kappa^2 (-)^{d^2}~^{\ast}J,}
\eqn\bianchi{dF \equiv 0,}
where the rank $d$ source $J$ is given by
\eqn\defj{J^{M_1 \ldots M_d} = T_d \int d^d \xi
\varepsilon^{i_1 i_2 \ldots i_d}
\partial_{i_1} X^{M_1} \partial_{i_2} X^{M_2} \ldots \partial_{i_d}
X^{M_d}{\delta^D (x-X)\over \sqrt{-g}}.}
Let us introduce the dual worldvolume dimension, $\tilde{d}$, by
\eqn\dtilde{\tilde{d} \equiv D - d - 2.}
We may now define two conserved charges:  the Noether ``electric'' charge
\eqn\elcharge{e_d = {1\over \sqrt{2}\kappa} \int\limits_{S^{\tilde{d}+1}}
e^{-\alpha (d)\phi}{}^{\ast}F,}
where $S^{\tilde{d}+1}$ is the ($\tilde{d}+1$)-sphere surrounding the
elementary
 ($d-1$)
brane, and the topological ``magnetic'' charge
\eqn\macharge{g_{\tilde{d}} = {1\over \sqrt{2} \kappa} \int\limits_{S^{d+1}}
F.}
This latter charge will be non-zero if the action $I_D$ admits a solitonic
$\tilde{d}$-dimensional extended object (a ``($\tilde{d}-1$)-brane'').  These
charges
obey a Dirac quantization condition \refs{\nep,\tei},
\eqn\charquan{{e_d g_{\tilde{d}}\over 4\pi} = {n\over 2}, \qquad n =
{\rm integer}}
analogous to the ($d=1, D=4$) condition that relates electric and magnetic
charges.  At
this stage, of course, it is not yet obvious that the system admits either
elementary or
solitonic extended object solutions, nor if they do, what are the values of
the
electric
and magnetic charges $e_d$ and $g_{\tilde{d}}$.

Let us first consider the field equations resulting from $I_D + S_d$.
The Einstein
equation is
\eqn\einstein{\eqalign{\sqrt{-g}&\Bigg[R^{MN} - {1\over 2} g^{MN} R - {1\over
2} (\partial^M \phi\partial^N \phi - {1\over 2} g^{MN} (\partial\phi)^2)\cr
&~-{1\over 2} {1\over d!} (F^M\,_{M_1 \ldots M_d} F^{NM_1 \ldots M_d} - {1\over
2(d+1)}
g^{MN} F^2) e^{-\alpha (d)\phi}\Bigg]\cr
&=\kappa^2 \sqrt{-g} T^{MN} ((d-1)-{\rm brane}),\cr}}
where the energy-momentum tensor is given by
\eqn\engmtensor{T^{MN} ((d-1)-{\rm brane}) = -T_d \int d^d \xi \sqrt{-\gamma}
 \gamma^{ij} \partial_i
X^M \partial_j X^N e^{\alpha\phi/d} {\delta^D (x-X)\over \sqrt{-g}},}
the antisymmetric tensor equation is
\eqn\antieq{\partial_M (\sqrt{-g} e^{-\alpha\phi} F^{M M_1 \ldots M_d}) =
 2\kappa^2 T_d \int d^d \xi
\varepsilon^{i_1 \ldots i_d} \partial_{i_1} X^{M_1} \ldots \partial_{i_d}
X^{M_d}
\delta^D (x-X),}
and the dilaton equation is
\eqn\dilaeq{\eqalign{&\partial_M (\sqrt{-g} g^{MN} \partial_N \phi)+
{\alpha (d)\over 2(d+1)!}\sqrt{-g} e^{-\alpha (d)\phi} F^2 \cr
&={\alpha (d) \kappa^2 T_d\over d}\int d^d \xi \sqrt{-\gamma} \gamma^{ij}
\partial_i X^M \partial_j X^N g_{MN} e^{\alpha (d)\phi/d} \delta^D (x-X).\cr}}
Furthermore, the ($d-1$)-brane field equations are
\eqn\braneq{\eqalign{\partial_i (\sqrt{-\gamma} \gamma^{ij} \partial_j X^N
g_{MN}e^{\alpha (d)\phi/d})&-{1\over 2} \sqrt{-\gamma} \gamma^{ij}
\partial_i X^N
\partial_j X^P\partial_M (g_{NP} e^{\alpha (d)\phi/d})\cr
&-{1\over d!} \varepsilon^{i_1 \ldots i_d} \partial_{i_1} X^{M_1} \ldots
\partial_{i_d} X^{M_d} F_{M M_1 \ldots M_d} = 0,\cr}}
and
\eqn\defgama{\gamma_{ij} = \partial_i X^M \partial_j X^N g_{MN}
e^{\alpha (d)\phi/d}.}

\newsec{\bf  The elementary ($d-1$)-brane}

To solve these coupled field-$(d-1)$-brane equations we begin by making an
ansatz
for the $D$-dimensional metric $g_{MN}$, $d$-form $A_{M_1} \ldots M_d$,
dilaton
$\phi$
and coordinates $X^M (\xi)$ corresponding to the most general $d/(D-d)$ split
invariant
under $P_d \times SO (D-d)$ where $P_d$ is the $d$-dimensional Poincar\'e
group.
  We
split the indices
\eqn\splitx{x^M = (x^{\mu}, y^m),}
where $\mu = 0, 1 \ldots (d-1)$ and $m = d, d+1, \ldots (D-1)$, and write the
line-element as
\eqn\supmetr{ds^2 = e^{2A} \eta_{\mu\nu} dx^{\mu} dx^{\nu} + e^{2B} \delta_{mn}
 dy^m dy^n,}
and the $d$-form gauge field as
\eqn\dform{A_{\mu_1 \ldots \mu_d} = - {1\over ^dg} \varepsilon_{\mu_1 \ldots
\mu_d} e^C,}
where $^dg$ is the determinant of $g_{\mu\nu}$, $\varepsilon_{\mu_1 \ldots
\mu_d} \equiv g_{\mu_1 \nu_1} \ldots g_{\mu_d \nu_d}
\varepsilon^{\nu_1 \ldots \nu_d}$ and
$\varepsilon^{012 \ldots (d-1)} = 1$ i.e. $A_{01 \ldots (d-1)} = - e^C$.  All
other
components of $A_{M_1 \ldots M_d}$ are set to zero.  $P_d$ invariance requires
that the
arbitrary functions A, B, C depend only on $y^m$; $SO(D-d)$ invariance then
requires that
this dependence be only through $y = \sqrt{\delta_{mn} y^m y^n}$.
Similarly our
 ansatz
for the dilaton is
\eqn\ansaph{\phi = \phi (y).}
In the ($d-1$)-brane sector we also split
\eqn\bransplit{X^M = (X^{\mu}, Y^m),}
and make the static gauge choice
\eqn\staticgauge{X^{\mu} = \xi^{\mu},}
and the ansatz
\eqn\ansatzy{Y_m = {\rm constant}.}
Substituting these ansatz into \defgama\ yields
\eqn\newgama{\gamma_{ij} = e^{2A + \alpha(d)\phi/d} \eta_{ij},}
and the only non-vanishing components of the field strength are
\eqn\fieldstr{F_{m\mu_1 \ldots \mu_d} = - {1\over ^dg} \varepsilon_{\mu_1
\ldots \mu_d} \partial_m e^C.}
Then the $\mu\nu$ components of the Einstein equation \einstein\ reduce to a
 single equation
\eqn\munucomp{\eqalign{&e^{(d-2) A + {\tilde d} B} \delta^{mn} \bigg[(d-1)
\partial_m \partial_n A +
{d(d-1)\over 2} \partial_m A \partial_n A + ({\tilde d} + 1)
\partial_m \partial_n B\cr
&+{({\tilde d} + 1) {\tilde d}\over 2} \partial_m B \partial_n B + \tilde d
(d-1) \partial_m A\partial_n B\cr
&+ {1\over 4} e^{-2dA + 2C - \alpha (d)\phi} \partial_m C \partial_n C +
{1\over 4} \partial_m \phi \partial_n \phi\bigg]\cr
&= - \kappa^2 T_d e^{(d-2) A + \alpha(d) \phi/2} \delta^{D-d} (y),\cr}}
and the $mn$ components reduce to
\eqn\mncomp{\eqalign{&e^{dA + (\tilde d - 2) B} \bigg[-\tilde d
\partial^m \partial^n B + \delta^{m n}\tilde d  \delta^{kl}
\partial_k \partial_l B\cr
&-d\partial^m \partial^n A + d \delta^{mn} \delta^{kl} \partial_k \partial_l A
 - d\partial^m A \partial^n A + {d(d+1)\over 2} \delta^{mn} \delta^{kl}
\partial_k A\partial_l A\cr
&+d(\partial^m A \partial^n B + \partial^m B \partial^n A + (\tilde d - 1)
\delta^{mn} \delta^{kl} \partial_k A \partial_l B)\cr
&-{1\over 2} \partial^m \phi \partial^n \phi + {1\over 4} \delta^{mn}
\delta^{kl} \partial_k \phi \partial_l \phi\bigg] \cr
&- {1\over 2} e^{-dA + (\tilde d - 2) B +
2C - \alpha (d)\phi}
\bigg[-\partial^m C \partial^n C + {1\over 2} \delta^{mn} \delta^{kl}
\partial_k C \partial_l C\bigg]\cr &=0.\cr}}
The antisymmetric tensor field equation \antieq\ becomes
\eqn\newanti{\delta^{mn} \partial_m \bigg[e^{-\alpha (d)\phi - dA +
\tilde d B}
\partial_n e^C\bigg] = 2\kappa^2 T_d \delta^{D-d} (y),}
and the dilaton equation \dilaeq\ becomes
\eqn\newdila{\eqalign{&\delta^{mn} \partial_m \bigg(e^{dA + \tilde d B}
\partial_n \phi\bigg) - {\alpha (d)\over 2}
e^{-dA + \tilde d B + 2C - \alpha (d)\phi} \delta^{mn} \partial_m C
\partial_n C\cr
&=\alpha (d) \kappa^2 T_d e^{dA + \alpha (d)\phi/2} \delta^{(D-d)} (y).\cr}}
Finally, the ($d-1$)-brane equation \braneq\ becomes
\eqn\newbrane{\partial_m (e^{dA + \alpha (d)\phi/2} - e^C) = 0.}
Hence we have five equations for the four unknown functions A, B, C, $\phi$ and
the
unknown parameter $\alpha (d)$.

The unique solution, assuming that $g_{MN}$ tends asymptotically to
$\eta_{MN}$,
 is given
by
\eqn\solution{\eqalign{A&={\tilde{d}\over 2(d+\tilde{d})} (C - C_o),\cr
B&= - {d\over 2(d+\tilde{d})} (C - C_o),\cr
{\alpha (d)\over 2} \phi&={\alpha^2 (d)\over 4} (C - C_o) + C_o,\cr}}
where $C_o = \alpha \phi_o/2$ and $\phi_o$ is the dilaton vev.
$C$ is given by
\eqn\cexpre{\eqalign{e^{-C}&=e^{-C_o} + {k_d\over y^{\tilde{d}}}, \qquad
\tilde{d} > 0\cr
&=e^{-C_o} - {\kappa^2 T_d\over \pi} ln~y, \qquad \tilde{d} = 0\cr}}
and
\eqn\defk{k_d = 2\kappa^2 T_d/\tilde{d}~\Omega_{\tilde{d}+1},}
where $\Omega_{\tilde{d}+1}$ is the volume of $S^{\tilde{d}+1}$.  The parameter
$\alpha (d)$ is
given by
\eqn\alphaexpre{\alpha^2 (d) = 4 - {2d\tilde{d}\over d + \tilde{d}}.}

Note, incidentally, that for these solutions, the coefficients of the
$\delta$-function in \munucomp\ and \newdila\ vanish at $y = 0$. So the
Einstein equation and the dilaton equation are essentially source-free;
only in the antisymmetric tensor equation is a $\delta$-function source.
We shall return to this in section 5.

A crucial result of this section is that we have fixed the constant
$\alpha (d)$
as in (3.18) by the requirement that our theory (2.1) yield elementary $(d -
1)$-brane solutions.

\newsec{\bf  Bogomol'nyi bounds and the ``no-force'' condition}

The mass per unit ($d-1$)-volume of the elementary ($d-1$)-brane is given by
\eqn\defmass{{\cal M}_d = \int d^{D-d} y \theta_{oo},}
where $\theta_{MN}$ is the total energy-momentum pseudotensor of the combined
gravity-matter system.  One may generalize the $d=2$, $\phi_o = 0$, arguments
of
Dabholkar et al \dabghr\ to arbitrary $d$ and non-vanishing $\phi_o$ and
establish a Bogolmol'nyi bound
\eqn\bogom{\kappa {\cal M}_d \geq {1\over \sqrt{2}} \mid e_d \mid e^{C_o} =
{1\over \sqrt{2}} \mid e_d \mid e^{\alpha (d)\phi_o/2},}
where $e_d$ is the electric charge of \elcharge.  For the solution of
Section 3, we find
\eqn\masexpre{{\cal M}_d = T_d e^{C_o}.}
To compute $e_d$ it is convenient to introduce polar coordinates
\eqn\polar{y^m = (y, \theta^i),}
where $i = 1, \ldots, (\tilde{d}+1)$, so that
\eqn\metricy{\delta_{mn} dy^m dy^n = dy^2 + y^2 d\Omega^2_{\tilde{d}+1},}
where $d\Omega^2_{\tilde{d}+1}$ is the metric on the unit $S^{\tilde{d}+1}$.
Then we note from \fieldstr\ that
\eqn\strexpre{F_{y \mu_1 \ldots \mu_d} = -{1\over ^dg} \varepsilon_{\mu_1
\ldots \mu_d} \partial_y e^C ,}
The dual of $F,~^{\ast}F$, has non-vanishing components only in the
$\theta^i$ directions
\eqn\dualf{\sqrt{-g} ^{\ast}F^{\theta_1 \ldots \theta_{D-d-1}} = -(-)^{(D-d)
(d+1)} e^{2C} \partial_y e^{-C},}
Hence, using (3.15--18) we find
\eqn\dualfa{e^{-\alpha\phi}{}^{\ast}F_{\theta_1 \ldots \theta_{D-d-1}} =
(-)^{(D-d) (d+1)} 2\kappa^2 T_d {\varepsilon_{\theta_1 \ldots \theta_{D-d-1}}
\over \Omega_{\tilde{d}+1}}.}
It follows from \elcharge\ that
\eqn\elchargea{e_d = \sqrt{2} \kappa T_d (-)^{(D-d) (d+1)},}
and hence
\eqn\massagain{{\cal M}_d = {1\over \sqrt{2}} \mid e_d \mid e^{\alpha (d)
\phi_o/2},}
and the bound in \bogom\ is saturated.  This shows that these elementary
($d-1$)-brane
solutions are stable.

So far we have concentrated on single ($d-1$)-brane solutions of the field
equations.
However, there is a straightforward generalization to exact, stable
 multi-($d-1$)-brane
configurations obtained by a linear superposition of the solutions \cexpre,
\eqn\multisol{e^{-C} = e^{-C_o} + \sum\limits_l {k_d\over \mid \underline{y}
 - \underline{y}_l \mid^2},}
where $y_l$ corresponds to the position of each ($d-1$)-brane.  The ability to
superpose
solutions of this kind is a well-known phenomenon in soliton and instanton
physics and
goes by the name of the ``no-force condition'' \dabghr.  In the present
context, it
 means
that the mutual gravitational-dilaton attraction of two  separated
($d-1$)-branes
is exactly cancelled by an equal and opposite contribution from the
antisymmetric
tensor.  To see this explicitly, consider the multi-($d-1$)-brane configuration
\multisol\ with, for example, n ($d - 1$)-branes as sources. In general, we do
not have the transverse $SO(D - d)$ symmetry, but we still have the $P_d$
Poincare symmetry for the configuration \multisol. Let each ($d - 1$)-brane
with
label $l$ satisfy $X^\mu (l) = \xi^\mu$ so that, in particular, they all have
the same orientation. The Lagrangian for each of the ($d - 1$)-branes with
label $l$ in the fields of the sources given by (3.1--4) is,
from \gepbact\
\eqn\noforce{{\cal L}_d = -T_d \Bigg[ \sqrt{-det (e^{2A + \alpha (d)\phi/d}
\eta_{ij} + e^{2B +\alpha (d)\phi/d} \partial_i Y^m (l) \partial_j Y_m (l)} -
e^C\Bigg]}
corresponding to a potential
\eqn\forcepoten{V = T_d (e^{dA + \alpha (d)\phi/2} - e^C),}
but this vanishes by \newbrane.  This generalizes to arbitrary $d$ and $D$ the
``no-force condition'' for strings \dabghr, fivebranes \duflfb\ in $D = 10$
and membranes in $D = 11$ \dufs.  Expanding out \noforce\ we find
\eqn\lagexpand{{\cal L} = - {T_d \over2}~e^{(d-2) A + 2B + \alpha (d)\phi/2}
\eta^{ij} \partial_i Y^m \partial_j Y_m +\ldots ,}
and so the absence of velocity-dependent forces corresponds to
\eqn\absencevdf{(d-2) A + 2B + \alpha (d)\phi/2 = {\rm constant},}
which is indeed satisfied by virtue of \solution\ and we find that the
constant is just $C_o$.  This generalizes to arbitrary $d$ and $D$, the absence
 of velocity dependent
forces for strings and fivebranes in $D = 10$ \calk.

\newsec{\bf  The solitonic ($\tilde{d}-1$)-brane}

The elementary ($d-1$)-branes we have discussed so far correspond to solutions
 of the
coupled field-brane system with action $I_D (d) + S_d$.  As such they exhibit
$\delta$-function singularities at $y = 0$.  They are characterized by a
non-vanishing
Noether ``electric'' charge $e_d$.  By contrast, we now wish to find solitonic
($\tilde{d}-1$)-brane, corresponding to solutions of the source free equations
 resulting
from $I_D (d)$ alone, which are regular at $y = 0$, and will be characterized
by a
non-vanishing topological ``magnetic'' charge $g_{\tilde{d}}$.  (Recall that
$\tilde{d} = D-d-2$).

To this end, we now make an ansatz invariant under $P_{\tilde{d}} \times
SO(D-\tilde{d})$.  Hence we write \splitx\ and \supmetr\ as before where now
$\mu = 0, 1 \ldots
(\tilde{d}-1)$ and $m = \tilde{d}, \tilde{d} + 1, \ldots (D-1)$.  The ansatz
for the
antisymmetric tensor, however, will now be made on the field strength rather
than on the
potential.  From section 3 we recall that a non-vanishing electric charge
corresponds to
\eqn\elchargeb{{1\over \sqrt{2} \kappa} e^{-\alpha\phi}{}^{\ast}F_{\tilde{d}+1}
 = e_d \varepsilon_{\tilde{d}+1}/\Omega_{\tilde{d}+1},}
where $\varepsilon_{\tilde{d}+1}$ is the volume form on $S^{\tilde{d}+1}$.
 Accordingly,
to obtain a non-vanishing magnetic charge, we make the ansatz
\eqn\ansazmag{{1\over \sqrt{2} \kappa} F_{d+1} = g_{\tilde{d}}
\varepsilon_{d+1}/\Omega_{d+1},}
where $\varepsilon_{d+1}$ is the volume form on $S^{d+1}$.  Since this is an
harmonic
form, $F$ can no longer be written globally as the curl of $A$, but it
satisfies the
Bianchi identities.  It is now not difficult to show that all the field
equations are
satisfied simply by making the replacement $d \rightarrow \tilde{d}$, and hence
$\alpha(d) \rightarrow \alpha (\tilde{d}) = - \alpha (d)$
in (3.15--18). For future reference we write the explicit solution
in the case $\phi_0 = 0$
\eqn\solisolution{\eqalign{ds^2 &= \bigg(1 + {k_{\tilde d}\over y^d}\bigg)^{
-d/(d + {\tilde d})} dx^\mu dx_\mu + \bigg(1 + {k_{\tilde d}\over y^d}\bigg)^{
\tilde d /(d + {\tilde d})} dy_m dy_m,\cr
e^{ 2\phi} & = \bigg(1 + {k_{\tilde d}\over y^d}\bigg)^{\alpha(d)},\cr
F_{d + 1} &= \sqrt {2} \kappa g_{\tilde d} \varepsilon_{d + 1}
/ \Omega_{d + 1}.\cr}}
Note that by this device, we have found solutions everywhere including
$y = 0$, since the $\delta$-functions were already absent in the Einstein
and dilaton equations.

It follows that the mass per unit ($\tilde{d}-1$)-volume now saturates a bound
involving
the magnetic charge
\eqn\magbogom{\eqalign{{\cal M}_{\tilde{d}}&={1\over \sqrt{2}} \mid g_{
\tilde{d}} \mid e^{\alpha (\tilde{d}) \phi_o/2}\cr
&={1\over \sqrt{2}} \mid g_{\tilde{d}} \mid e^{-\alpha (d) \phi_o/2}.\cr}}
Note that the $\phi_o$ dependence is such that ${\cal M}_{\tilde{d}}$ is large
for small ${\cal M}_d$ and vice-versa.

The electric charge of the elementary solution and the magnetic charge of
the soliton
solution obey a Dirac quantization rule \refs{\nep,\tei}
\eqn\pbdirac{e_d g_{\tilde{d}} = 2 \pi n, \qquad n = {\rm integer},}
and hence from \elchargea\
\eqn\machargea{(-)^{(D-d)(d+1)} \tilde{g}_{\tilde{d}} = 2\pi n/\sqrt{2}
\kappa T_d,}

\newsec{\bf  Duality}

We now wish to consider the theory ``dual'' to \lgeact\ for which the roles of
 field
equations \fieldeq\ and Bianchi identities \bianchi\ are interchanged.  To this
end let us write the action
\eqn\dualact{\tilde{I}_D (\tilde{d}) = {1\over 2 \kappa^2} \int d^Dx \sqrt{-g}
 \Bigg(R - {1\over 2}
(\partial\phi)^2 - {1\over 2(\tilde{d}+1)!} e^{\alpha (d)\phi}
\tilde{F}_{\tilde{d}+1}^2\Bigg),}
where the rank ($\tilde{d}+1$) field strength $\tilde{F}$ is given by
\eqn\dualstr{\tilde{F}_{\tilde{d}+1} = d \tilde{A}_{\tilde{d}},}
$\alpha (d)$ is the same constant as appearing in \lgeact\ but appears with
opposite sign, i.e
\eqn\dualpha{\alpha(\tilde d ) = - \alpha (d).}
  Allow
these fields to couple to an elementary $\tilde{d}$-dimensional extended object
(a``($\tilde{d}-1$)-brane'') with action
\eqn\dualpbact{\eqalign{\tilde{S}_{\tilde{d}} = T_{\tilde{d}} \int d^{\tilde{d}
} \xi \Bigg(&-{1\over 2}
\sqrt{-\gamma} \gamma^{ij} \partial_i X^M \partial_j X^N g_{MN}
 e^{-\alpha (d)\phi/
\tilde{d}} + {(\tilde{d}-2)\over 2} \sqrt{-\gamma}\cr
&-{1\over \tilde{d}!} \varepsilon^{i_1 i_2 \ldots i_{\tilde{d}}} \partial_i X^{
M_1}\partial_{i_2} X^{M_2} \ldots \partial_{i_d} X^{M_{\tilde{d}}} \tilde{A}_{
M_1 M_2 \ldots M_{\tilde{d}}}\bigg).\cr}}
The $\phi$ dependence is such that under the rescaling
\eqn\dualscaling{\eqalign{g_{MN}&\rightarrow \tilde{\lambda}^{2\tilde{d}/(D-2)}
 g_{MN},\cr
\tilde{A}_{M_1 \ldots M_{\tilde{d}}}&\rightarrow \tilde{\lambda}^d \tilde{A}_{
M_1 \ldots M_{\tilde{d}}},\cr
e^{\phi}&\rightarrow \tilde{\lambda}^{-2\tilde{d} (D-\tilde{d}-2)/(D-2)
\alpha (d)} e^{\phi},\cr
\gamma_{ij}&\rightarrow \tilde{\lambda}^2 \gamma_{ij},\cr}}
both actions scale the same way
\eqn\dualactscal{\eqalign{\tilde{I}_D (\tilde{d})&\rightarrow \tilde{\lambda}^{
\tilde{d}} I_D (d),\cr
\tilde{S}_{\tilde{d}}&\rightarrow \tilde{\lambda}^{\tilde{d}}
\tilde{S}_{\tilde{d}}.\cr}}
The field equations and Bianchi identities of the $\tilde{A}$ field may be
written
\eqn\dufieldeq{d^{\ast} (e^{\alpha (d)\phi} \tilde{F})=
2\kappa^2 (-)^{\tilde{d}^2}{}^{\ast}\tilde{J},}
\eqn\dubianchi{d\tilde{F} = 0.}
It should be clear that the system described by $\tilde{I}_D (\tilde{d}) +
\tilde{S}_{\tilde{d}}$ admit the same elementary solutions as that described by
$I_D (d)+ S_d$ and that $\tilde{I}_D (\tilde{d})$ alone admits the same
solitonic solutions as
$I_D (d)$ alone, provided we everywhere make the replacement $d \rightarrow
\tilde{d}$
and hence $\alpha (d) \rightarrow \alpha (\tilde{d}) = - \alpha (d)$.  In
particular the
Noether electric charge is given by
\eqn\duelcharge{\tilde{e}_{\tilde{d}} = {1\over \sqrt{2}\kappa} \int\limits_{
S^{d+1}} e^{\alpha\phi}{}^{\ast}\tilde{F}_{d+1},}
and the topological magnetic charge by
\eqn\dumacharge{\tilde{g}_d = {1\over \sqrt{2} \kappa} \int\limits_{S^{
\tilde{d}+1}}\tilde{F}_{\tilde{d}+1},}
and they obey the condition
\eqn\dualdirac{{\tilde e}_{\tilde{d}} {\tilde g}_d = 2\pi n.}

So far we have discovered that the equations of $I_D (d)$ admit an elementary
($d-1$)-brane solution and a solitonic ($\tilde{d}-1$)-brane solution.
Conversely, the
equations of $\tilde{I}_D (\tilde{d})$ admit an elementary
($\tilde{d}-1$)-brane
 solution
and a solitonic ($d-1$)-brane solution.  We now wish to go a step further and
assert that
the ($d-1$) brane is ``dual'' to the ($\tilde{d}-1$)-brane.  In its strongest
sense this
means that the two theories are equivalent descriptions of the same physics.
In the
present context, however, we simply make the assumption that the $I_D (d)$ and
$\tilde{I}_D (\tilde{d})$ are equivalent i.e we assume that the metric $g_{MN}$
and
dilaton $\phi$ are the same and that the ($\tilde{d}+1$)-form field strength
$\tilde{F}_{\tilde{d}+1}$ is dual to the ($d+1$)-form field strength $F_{d+1}$.
 More
precisely,
\eqn\duality{\tilde{F}_{\tilde{d}+1} = e^{-\alpha (d)\phi}{}^{\ast}F_{d+1}}
so that the (source-free) field equations and Bianchi identities of $I_D (d)$,
\fieldeq\ and \bianchi, become the Bianchi identities and (source-free) field
equations of $\tilde{I}_D (\tilde{d})$, \dubianchi\ and \dufieldeq.  This leads
 immediately to
\eqn\charelation{\eqalign{e_d&=\tilde{g}_d,\cr
g_{\tilde{d}}&=\tilde{e}_{\tilde{d}},\cr}}
and hence
\eqn\tensiondirac{\kappa^2 T_d T_{\tilde{d}} =  |n| \pi}

The duality assumption also leads to a relation between the dimensionless loop
expansion
parameters of the ($d-1$)-brane and the ($\tilde{d}-1$)-brane.  To see this we
note that
metrics appearing naturally in ($d-1$)-brane and ($\tilde{d}-1$)-brane
$\sigma$-models \gepbact\ and \dualpbact\ are
\eqn\sigmacano{g_{MN} (d)=e^{\alpha (d)\phi/d} g_{MN} ({\rm canonical}),}
\eqn\dusigmac{g_{MN} (\tilde{d})=e^{-\alpha (d)\phi/\tilde{d}} g_{MN} ({\rm
canonical}).}
If we rewrite $I_D (d)$ and $\tilde{I}_D (\tilde{d})$ in these variables we
find
\eqn\sigmact{\eqalign{I_D(d)&={1\over 2\kappa^2} \int d^Dx \sqrt{-g} e^{-(D-2)
\alpha (d)\phi/2d} \Bigg[R\cr
&\quad -{1\over 2} \bigg(1 - {\alpha^2 (D-1) (D-2)\over 2d^2}\bigg) (\partial
\phi)^2 - {1\over 2 \cdot (d+1)!} F^2_{d+1}\Bigg],\cr}}
and
\eqn\dusigmact{\eqalign{\tilde{I}_D (\tilde{d})&={1\over 2\kappa^2} \int d^Dx
\sqrt{-g} e^{(D-2)\alpha (d)\phi/2\tilde{d}} \Bigg[R\cr
&\quad -{1\over 2} \bigg(1 - {\alpha^2 (D-1) (D-2)\over 2\tilde{d}^2}\bigg)
(\partial\phi)^2 -
{1\over 2 (\tilde{d}+1)!} \tilde{F}^2_{\tilde{d}+1}\Bigg].\cr}}
Note that in both cases a common dilaton-dependent factor appears.  This
reveals that the
($d-1$)-brane loop counting parameter is
\eqn\looparam{{\rm g}_d = e^{(D-2) \alpha (d) \phi_o/4d},}
and the ($\tilde{d}-1$)-brane loop counting parameter is
\eqn\dulooparam{{\rm g}_{\tilde{d}} = e^{-(D-2) \alpha (d)\phi_o/4\tilde{d}}.}
Hence
\eqn\loopduality{{\rm g}_d^d = 1/{\rm g}_{\tilde{d}}^{\tilde{d}},}
and strongly coupled ($d-1$) branes correspond to weakly coupled
($\tilde{d}-1$)
 branes
and vice-versa.

Finally we note that, in the case of $d = 2$, the following field redefinition
\eqn\fieldredef{(D-2) \alpha (2) \phi = 8 \Phi}
yields from \sigmact\ an $I_D (d)$ which is $D$-independent, namely
\eqn\stringact{I_D (2) = {1\over 2\kappa^2} \int d^Dx \sqrt{-g} e^{-2\Phi}
\Bigg[R + 4(\partial\Phi)^2 - {1\over 2.3!} F_3^2\Bigg].}
This is a well-known result in string theory.  Curiously, there is no field
redefinition which renders $I_D (d)$ independent of $D$ for $d \not= 2$.
However, we may
dualize \stringact\ to obtain
\eqn\dustringact{\tilde{I}_D (D-4) = {1\over 2\kappa^2} \int d^Dx \sqrt{-g}~e^{
4\Phi/D-4} \Bigg[R
- {4 (D-10)\over  (D-4)^2} (\partial\Phi)^2 - {1\over 2(D-3)!}
\tilde{F}^2_{D-3}\Bigg].}
In these string variables the metric of the elementary string is given by
\eqn\stringmetric{ds^2 = \bigg(1 + {k_2 e^{C_o}\over y^{D-4}}\bigg)^{-1} \eta_{
\mu\nu} dx^{\mu} dx^{\nu} + \delta_{mn} dy^m dy^n}
with $\mu = 0,1$ and $m = 1 \ldots D-2$.  Also
\eqn\stringalpha{\alpha (2) = \sqrt{{8\over D-2}},}
so
\eqn\redefphi{\Phi = {1\over 2} (C - C_o) + {D-2\over 4} C_o ,}
where
\eqn\stringc{\eqalign{e^{-C}&=e^{-C_o} + {k_2\over y^{D-4}}, \qquad D > 4 \cr
&=e^{-C_o} - {\kappa^2 T_2\over \pi}~ln~y. \qquad D = 4 \cr}}

On the other hand the solitonic ($D-5$)-brane is given by
\eqn\dustringmetr{ds^2 = \eta_{\mu\nu} dx^{\mu} dx^{\nu} + \bigg(1 + {k_{D-4}
\over y^2} e^{C_o}\bigg)\delta_{mn} dy^m dy^n ,}
where $\mu = 0 \ldots D-5$ and $m = D - 4, \ldots, D - 1$.  Also
\eqn\dualpha{\alpha (D-4) = - \sqrt{{8\over D-2}},}
so
\eqn\dualphi{\Phi = - {1\over 2}~(C - C_o) - {(D-2)\over 4} C_o ,}
where
\eqn\dualc{e^{-C} = e^{-C_o} + {k_{D-4}\over y^2}.}
We note that in these string $\sigma$-model variables the transverse part of
 the
 metric
in \stringmetric\ is flat and the spacetime part of the metric in
\dustringmetr\
 is flat.  These
 are
therefore free field theories from the point of view of conformal field theory.

\newsec{\bf  Higher loops}

$D$-dimensional strings involve two kinds of loop expansion:  quantum $D = 10$
strings
loops ($L$) with loop expansion parameter
${\kappa^2 e^{2\phi}\over (2\pi)^{D/2}}$
and classical $d = 2$ $\sigma$-model loops with loop expansion parameter
$\alpha'_2 \equiv 1/2\pi T_2$, assuming we use the string $\sigma$-model
metric.  Similarly a $D$-dimensional
($d-1$)-brane will presumably require quantum $D$-dimensional ($d-1$)-brane
loops ($L$) with loop expansion parameter
${\kappa^2 e^{(D-2) \alpha (d) \phi/2d}\over (2\pi)^{D/2}}$ and
classical $d$-dimensional $\sigma$-model loops with loop expansion parameter
$\alpha'_d \equiv 1/(2\pi)^{d/2} T_d$.  Let us consider the purely
gravitational contribution to the
resulting effective action, using the ($d-1$)-brane $\sigma$-model metric:
\eqn\graveffecact{{\cal L}_{LL+m} = a_{LL+m} {1\over 2\kappa^2} \sqrt{-g} e^{-
(D-2) \alpha (d)\phi/2d}
\Bigg({2\kappa^2 e^{(D-2) \alpha (d)\phi/2d}\over (2\pi)^{D/2}}\Bigg)^L
\alpha'_d{}^m R^n ,}
where $R^n$ is symbolic for a scalar contribution of $n$ Riemann tensors each
of
dimension 2.  One could also include covariant derivatives of $R$ but, for our
purposes \graveffecact\ will be sufficient.  The $a_{LL+m}$ are numerical
coefficients, not involving $\pi$. Since
\eqn\cannodim{[{\cal L}_{LL+m}] = D, \qquad [\kappa^2] = 2 - D, \qquad
[\alpha'_d] = -d,}
we have, on dimensional grounds,
\eqn\looprelation{dm = 2 (n - 1) - (D - 2) L.}

By the same argument, a $D$-dimensional ($\tilde{d}-1$)-brane will require
quantum
$D$-dimensional ($\tilde{d}-1$)-brane loops ($\tilde{L}$) with loop expansion
parameter
${\kappa^2 e^{-(D-2) \alpha (d)\phi/2\tilde{d}}\over (2\pi)^{D/2}}$ and
classical
$\tilde{d}$-dimensional $\sigma$-model loops with loop expansion parameter
$\alpha'_{\tilde{d}} \equiv 1/(2\pi)^{\tilde{d}/2} T_{\tilde{d}}$.  The
corresponding
Lagrangian using the ($\tilde{d}-1$)-brane $\sigma$-model metric is
\eqn\dugraeffec{\tilde{{\cal L}}_{\tilde{L}+\tilde{m} \tilde{L}} = \tilde{a}_{
\tilde{L}+ \tilde {m} L}
{1\over 2\kappa^2} \sqrt{-g} e^{(D-2) \alpha (d)\phi/2\tilde{d}}
\bigg({2\kappa^2 e^{-(D-2) \alpha (d)\phi/2\tilde{d}}\over (2\pi)^{D/2}}
\bigg)^{\tilde{L}} \alpha'_{\tilde{d}}{}^{\tilde{m}} R^n .}
Again, on dimensional grounds,
\eqn\duloprelat{\tilde{d} \tilde{m} = 2 (n - 1) - (D - 2) \tilde{L}.}

Our fundamental assumption is that ${\cal L}$ and $\tilde{\cal L}$ are related
by duality
which implies, in particular, that the purely gravitational contributions
should be
identical when written in the same variables.  So transforming to the canonical
metric
using \sigmacano\ and \dusigmac, we find
\eqn\canoact{{\cal L}_{LL+m} = {1\over 2\kappa^2} a_{LL+m} \bigg({2\kappa^2
\over (2\pi)^{D/2}}\bigg)^L \alpha'_d{}^m e^{-\alpha(d) m \phi/2}
\sqrt{-g} R^n,}
\eqn\ducanoact{\tilde{{\cal L}}_{\tilde{L}+\tilde{m} \tilde{L}} = {1\over
2\kappa^2}\tilde{a}_{\tilde{L}+\tilde{m}\tilde{L}} \bigg({2\kappa^2\over
(2\pi)^{D/2}}\bigg)^{\tilde{L}} \alpha'_{\tilde{d}}{}^{\tilde{m}}
e^{\alpha (d) \tilde{m} \phi/2} \sqrt{-g} R^n ,}
where we have dropped the terms like
$(\partial\phi)^{2m} R^{n-m}$~for~$m = 1, 2, \ldots, n$.  Bearing in mind that
 from \tensiondirac\ with unit integer,
\eqn\alphadirac{2\kappa^2 = (2 \pi)^{D/2} \alpha'_d \alpha'_{\tilde{d}},}
we find that ${\cal L}$ and $\tilde{{\cal L}}$ do coincide provided
\eqn\mtildem{m + \tilde{m} = 0,}
i.e from \looprelation\ and \duloprelat, provided
\eqn\lmrelation{m = \tilde{L} - L = - \tilde{m},}
\eqn\nlrelation{2n = \tilde{d} L + d \tilde{L} + 2,}
with
\eqn\atildea{a_{L\tilde{L}} = (-)^{(L - \tilde{L}) D} \tilde{a}_{L \tilde{L}},}
and hence
\eqn\resultact{{\cal L}_{L\tilde{L}} = a_{L\tilde{L}} (-)^{D L}
{1\over 2\kappa^2} \alpha'_d{}^{\tilde{L}}
\alpha'_{\tilde{d}}{}^L e^{\alpha (L-\tilde{L}) \phi/2} \sqrt{-g}
R^{(\tilde{d}L + d\tilde{L} +2)/2}.}
This generalizes the result of \dufllo, where \resultact\ was obtained for
 $d = 2, \tilde{d} = 6$.  Interestingly, it was there obtained in the context
of
the heterotic string, but here
we see that the result is, in fact, universal.  A similar analysis for the pure
antisymmetric tensor terms yields, in the canonical metric,
\eqn\tensoract{{\cal L}_{LL+m} = {1\over 2\kappa^2} a_{LL+m} \bigg({2\kappa^2
\over (2\pi)^{D/2}}\bigg)^L
\alpha'_d{}^m e^{-\alpha (d) m \phi/2} \sqrt{-g}\bigg({1\over 2 \cdot (d+1)!}
F_{d+1}^2 e^{-\alpha (d)\phi}\bigg)^n,}
\eqn\dutensoract{\tilde{{\cal L}}_{\tilde{L}+\tilde{m} L} = {1\over
2\kappa^2} \tilde{a}_{\tilde{L}+\tilde{m} \tilde{L}} \bigg({2\kappa^2\over
(2\pi)^{D/2}}\bigg)^{\tilde{L}} \alpha'_{\tilde{d}}{}^{\tilde{m}}
e^{\alpha (d) \tilde{m} \phi/2}
\sqrt{-g} \bigg({1\over 2 \cdot (\tilde{d}+1)!} F_{\tilde{d}+1}^2
e^{\alpha (d) \phi}\bigg)^n ,}
and again we find from \duality\ that the Bianchi identities and antisymmetric
 tensor field
equations of ${\cal L}$ and $\tilde{{\cal L}}$ are interchanged provided
(7.9--12) are
satisfied.

We note that under the rescalings \resclaw\ and \dualscaling\
${\cal L}_{L\tilde{L}}$ scales as
\eqn\actscaling{{\cal L}_{L\tilde{L}} \rightarrow \lambda^{d(1-\tilde{L})}
\tilde{\lambda}^{\tilde{d}(1-L)} {\cal L}_{L\tilde{L}}.}

As in \dufllo, \resultact\ gives rise to an infinite number of
non-renormalization theorems.  The
first of which is the absence of a cosmological term $\sqrt{-g} R^o$, assuming
that the
total Lagrangian is given by
\eqn\totalact{{\cal L} = \sum\limits_{L=0}^{\infty} \sum\limits_{\tilde{L}=0}^{
\infty} {\cal L}_{L\tilde{L}}.}
The second states that $\sqrt{-g} R$ appears only at ($L = 0, \tilde{L} = 0$)
and hence
the tree level action \lgeact\ does not get renormalized.

All this assumes, of course, that both the ($d-1$)-brane and the
($\tilde{d}-1$)-brane
are quantum mechanically consistent.  This will not be true in general but only
some
specific choices of $d$ and $D$.  We intend to return to the question of
quantum consistency elsewhere.

\newsec{\bf  $D = 10$}

So far, our analysis has kept both the dimension of the worldvolume, $d$, and
the dimension of spacetime, $D$, arbitrary.  However, in the case of strings
($d = 2$) we
know that certain spacetime dimensions are singled out for special attention.
 For
example, Green-Schwarz superstrings exist classically only for $D = 3, 4, 6$
and
 $10$ and, of these, only the $D = 10$ string is allowed
quantum-mechanically in
the sense
of being anomaly-free \gresw.  (By the way, ``$D = 10$'' is a loose way of
speaking about
central charge $c = 15$, so it could equally well mean a lower dimensional
string with
the correct amount of internal degrees of freedom).  Similarly, the critical
dimension of
the bosonic string is $D = 26$. Thus from our general discussion of sections
3 and 4, we see that, in addition to the elementary string solution \dabghr,
the bosonic string in $D = 26$ also admits a solitonic $21$-brane solution.

In the case of $N = 1$ super ($d-1$)-branes, the classically allowed
supersymmetric
extended objects have been classified by Achucarro et al \achetw\ and
correspond to
the
circles on the ``brane-scan'' of \duflbs.  However, this classification is
inadequate for
$N = 2$ super ($d-1$)-branes when $d > 2$ since such Type II supersymmetric
extended
objects require spin $> 1/2$ fields on the worldvolume
\refs{\calhsone,\calhstwo}, which were excluded
 by
the assumption in \achetw.  We have shown elsewhere \duflbs\ that the
($d-1$)-brane solutions of
sections 2 and 3 provide solutions of Type IIA supergravity in $D = 10$ for
$d = 1, 2, 3, 5, 6, 7$ (i.e $\tilde{d} = 7, 6, 5, 3, 2, 1$) only, and of Type
IIB supergravity in $D = 10$
 for $d = 2, 4, 6$ (i.e $\tilde{d} = 6, 4, 2$) only.
  The existence of the Type IIA
and
IIB superstring solutions was established by Dabholkar et al \dabghr, the Type
IIA and IIB
superfivebrane solutions by Callan et al \refs{\calhsone\calhstwo}, and the
self-dual Type
IIB superthreebrane
by the present authors \dufltb.  Now Horowitz and Strominger \hors\ have
exhibited a
two-parameter family of solutions of $D = 10$ Type IIA and B supergravity with
event
horizons: for $d = 1, 2, 3, 4, 5, 6, 7$ ``black ($d-1$)-branes''.  In some
respects, these
solutions resemble the Reissner-Nordstrom black-hole solution of general
relativity which
is known to admit unbroken supersymmetry in the extreme charge = mass limit.
Horowitz and
Strominger then conjectured that, in this limit, their black p-branes would
also
 be
supersymmetric and hence that there exist Type II super ($d-1$)-branes for all
these
values of $d$. As we have shown in \duflbs, this is indeed the case.

For $D = 10$, $\tilde{d} = 8 - d$ and hence
\eqn\tenalpha{\alpha (d) = {(4-d)\over 2} (-)^d.}
First of all, then, our generic Lagrangian \lgeact\ correctly describes the
bosonic
 sector
of the three-form field strength version of $N = 1$, $D = 10$ supergravity
\refs{\berddv,\cham}, where
\eqn\threeformv{I_{10} (2) = {1\over 2\kappa^2} \int d^{10}x \sqrt{-g} \bigg(R
 - {1\over 2} (\partial\phi)^2 - {1\over 2.3!} e^{-\phi} F_3^2\bigg)}
since $\alpha (2) = 1$.  The elementary solution of section 2 is therefore
given by
\eqn\tenabphi{\eqalign{A&={3\over 8} (C - C_o),\cr
B&=-{1\over 8} (C - C_o),\cr
\phi&={1\over 2} (C - C_o) + 2 C_o,\cr}}
where $C_o = \phi_o/2$,
\eqn\tenc{e^{-C} = e^{-C_o} + {k_2\over y^6},}
and
\eqn\ktwo{k_2 = \kappa^2 T_2/3 \Omega_7.}
This is the $D = 10$ string solution of Dabholkar et al \dabghr, generalized
to
non-vanishing $\phi_o$ \duflfb.

The generic Lagrangian \lgeact\ also describes the bosonic sector of the
seven-form
 field
strength version of $N = 1$, $D = 10$ supergravity \cha, which is dual to
\threeformv,
 namely
\eqn\sevenformv{I_{10} (6) = {1\over 2\kappa^2} \int d^{10}x \sqrt{-g} \bigg(R
 - {1\over 2} (\partial\phi)^2 - {e^{\phi}\over 2.7!} F_7^2\bigg),}
since $\alpha (6) = -1$.  The elementary solution of section 2 is therefore
\eqn\dutenabphi{\eqalign{A&={1\over 8} (C - C_o),\cr
B&=-{3\over 8} (C - C_o),\cr
\phi&=-{1\over 2} (C - C_o) - 2 C_o,\cr}}
where $C_o = - \phi_o/2$,
\eqn\dutenc{e^{-C} = e^{-C_o} + {k_6\over y^2},}
and
\eqn\ksix{k_6 = \kappa^2 T_6/\Omega_3,}
This is the $D = 10$ fivebrane solution of Duff and Lu \duflfb.  The above
solutions are
dual with the elementary solutions of $I_{10} (2)$ corresponding to the
solitonic
solution of $I_{10} (6)$, and vice versa.  From \tensiondirac\ and
\loopduality, the tensions
obey
\eqn\tentensiondirac{\kappa^2 T_2 T_6 = |n|\pi,}
and the loop coupling constants obey
\eqn\tenloopdu{{\rm g}_2^2 {\rm g}_6^6 = 1,}
in agreement with \dufldu.  The mass per unit length of the
string is given by
\eqn\masstwo{{\cal M}_2 = T_2 e^{\phi_o/2},}
and the mass per unit five-volume of the fivebrane by
\eqn\masssix{{\cal M}_6 = T_6 e^{-\phi_o/2},}
As expected, the string gets heavier for weak fivebrane coupling and the
fivebrane gets
heavier for weak string coupling.

As shown in \dabghr and \duflfb, both these solutions preserve half the
spacetime supersymmetry.

Let us now turn our attention to $D = 10$ Type IIA supergravity whose action
is
given,
in canonical variables, by
\eqn\tentypeiiact{\eqalign{I_{10} (IIA) = {1\over 2\kappa^2} \int&d^{10}x
\sqrt{-g} \bigg[R - {1\over 2}
(\partial\phi)^2 - {1\over 2.3!} e^{-\phi} F_3^2\cr
&-{e^{3\phi/2}\over 2.2!} F_2^2 - {1\over 2.4!} e^{\phi/2} F'_4\,^2\bigg] -
{1\over 8 \kappa^2}\int F_4 \wedge F_4 \wedge A_2 ,\cr}}
where $F'_4 = dA_3 + A_1 \wedge F_3$ comparison with \tenalpha\ shows that the
kinetic terms for gravity, dilaton and antisymmetric tensors are correctly
described by
the generic action $I_{10} (d)$ with $d = 1, 2, 3$ (i.e $\tilde{d} = 7, 6, 5$).
 Both
the elementary and solitonic $N = 1$ string and fivebrane solutions described
above
continue to provide solutions to Type IIA supergravity, as may be seen by
setting $F_2 =F_4 = 0$.  Again each preserve half the spacetime supersymmetry.
  (This establishes the
existence of Type IIA superfivebranes in $D = 10$, and these we surmise to be
dual, in
the sense of section 5, to Type IIA superstrings.)  This observation is not as
obvious
as it may seem in the case of elementary fivebranes or solitonic strings,
however,
since it assumes that one may dualize $F_3$.  Now the Type IIA action follows
by
dimensional reduction from the action of $D = 11$ supergravity, discussed in
the
 next
section.  There exists no dual formulation of this action \nictv, in which
$F_4$ is
replaced
by $F_7$, essentially because $A_3$ appears explicitly in the Chern-Simons term
$\int F_4 \wedge F_4 \wedge A_3$.  Since the $F_4$ and $F_3$ in $D = 10$
originate from $F_4$
in $D = 11$, this means that one cannot {\it simultaneously} dualize $F_3$ and
$F_4$ but one may do either {\it separately}.{\footnote\dg{We are grateful to
H.
Nishino for this observation.}}  By partial integration one may choose to have
 no
explicit $A_3$ dependence in the Chern-Simons term of \tentypeiiact\ or no
explicit
$A_2$
dependence, but not both.

By setting $F_2 = F_3 = 0$, we find elementary membrane ($d = 3$) and solitonic
fourbrane ($\tilde{d} = 5$) solutions, and then by dualizing $F_4$, elementary
fourbrane ($d = 5$) and solitonic membrane ($\tilde{d} = 3$) solutions.  From
(3.15--18) and \tenalpha, the elementary membrane solution is
given
by
\eqn\membransol{\eqalign{A&={5\over 16} (C - C_o),\cr
B&=-{3\over 16} (C - C_o),\cr
\phi&=-{1\over 4} (C - C_o) - 4 C_o,\cr}}
where $C_o = - \phi_o/4$,
\eqn\twoc{e^{-C} = e^{-C_o} + {k_3\over y^5},}
and
\eqn\kthree{k_3 = 2 \kappa^2 T_3/5\Omega_6 .}
The fourbrane solution is given explicitly by
\eqn\foursol{\eqalign{A&={3\over 16} (C - C_o),\cr
B&=-{5\over 16} (C - C_o),\cr
\phi&={1\over 4} (C - C_o) + 4 C_o ,\cr}}
where $C_o = \phi_o/4 ,$
\eqn\fourc{e^{-C} = e^{-C_o} + {k_5\over y^3},}
and
\eqn\kfive{k_5 = 2 \kappa^2 T_5/3\Omega_4 ,}
{}From \tensiondirac\ and \loopduality, the tensions obey
\eqn\thrtensiondirac{\kappa^2 T_3 T_5 = |n|\pi ,}
and the loop coupling constants obey
\eqn\threeloop{{\rm g}_3^3 {\rm g}_5^5 = 1,}
The mass per unit area of the membrane is given by
\eqn\massthree{{\cal M}_3 = T_3 e^{-\phi_o/4},}
and the mass per unit four-volume of the fourbrane by
\eqn\massfour{{\cal M}_5 = T_5 e^{\phi_o/4},}
Once again,  these membrane and fourbrane solutions break one
half of
the spacetime supersymmetries \duflbs\ and hence  there exist Type IIA
supermembranes
 and
Type IIA superfourbranes in accordance with a conjecture of Horowitz and
Strominger \hors, which we again expect to be dual to one another.

By setting $F_3 = F_4 = 0$, we find elementary particle ($d = 1$) and solitonic
sixbrane ($\tilde{d} = 7$) solutions, and then by dualizing $F_2$, elementary
 sixbrane
($d = 7$) and solitonic particle ($\tilde{d} = 1$) solutions.
 From (3.15--18) and
\tenalpha, the particle solution is given explicitly by
\eqn\tenparticsol{\eqalign{A&={7\over 16} (C - C_o),\cr
B&=-{1\over 16} (C - C_o),\cr
\phi&=-{3\over 4} (C - C_o) - {4\over 3} C_o ,\cr}}
where $C_o = - 3\phi_o/4 ,$
\eqn\particlec{e^{-C} = e^{-C_o} + {k_1\over y^7},}
and
\eqn\kone{k_1 = 2 \kappa^2 T_1/7\Omega_8 ,}
The sixbrane solution is given by
\eqn\sixbransol{\eqalign{A&={1\over 16} (C - C_o),\cr
B&=-{7\over 16} (C - C_o),\cr
\phi&={3\over 4} (C - C_o) + {4\over 3} C_o,\cr}}
where $C_o = 3\phi_o/4 ,$
\eqn\sevenc{e^{-C} = e^{-C_o} + {k_7\over y},}
and
\eqn\kseven{k_7 = 2\kappa^2 T_7/\Omega_2 ,}
{}From \tensiondirac\ and \loopduality, the tensions obey
\eqn\partitensiondirac{\kappa^2 T_1 T_7 = |n|\pi ,}
and the loop coupling constants obey
\eqn\particleloop{{\rm g}_1 {\rm g}_7^7 = 1,}
The mass of the particle is given by
\eqn\massone{{\cal M}_1 = T_1 e^{-3\phi_o/4},}
and the mass per unit six-volume of the sixbrane by
\eqn\massseven{{\cal M}_7 = T_7 e^{3\phi_o/4},}
These particle and sixbrane solutions break one half of the spacetime
supersymmetries and hence there exist Type II A superparticle and Type II A
supersixbrane in accordance with the conjecture of Horowitz and Strominger
\hors, which we once more expect to be dual to one another.

Let us now turn our attention to Type IIB supergravity in $D = 10$ whose
bosonic
 sector
consists of the graviton $g_{MN}$, a complex scalar $\phi$, a complex 2-form
$A_2$ (i.e $d = 2$ or, by duality, $d = 6$) and a real 4-form $A_4$ (i.e.
$d = 4$ which in $D =10$ is self-dual!).  Because of this self-duality of the
5-form field-strength $F_5$,
there exists no covariant action principle of the kind \lgeact\ and, strictly
speaking,
our previous analysis ceases to apply.  Nevertheless, we can apply the same
logic to
the equations of motion \sch\ and we find that the solutions again fall
into the
generic
category given by (3.15--18).  First of all, by truncation it is
 easy to see that
 the
same string and fivebrane of $N = 1$ supergravity continue to solve the field
equations
of Type IIB.  Moreover, they continue to break half the supersymmetries (but
there are
now twice as many since we start with $N = 2$ in $D = 10$ rather than $N = 1$).
 Hence
there exists Type IIB superstrings and superfivebranes, which are presumably
dual.

On the other hand, if we set to zero the three form $F_3$ and make the ansatz
(3.1--7) for the graviton, dilaton and $A_4$, setting to zero all
other independent components
of $F_5$ i.e those not related by the self-duality condition
\eqn\selfduf{F_5 = - ^{\ast}F_5,}
then we find the special case of (3.1--7) given by
$d = \tilde{d} = 4$ and hence
$\alpha= 0$ with $\phi =$ constant.  Explicitly, this self-dual threebrane is
 given by
\eqn\threesol{\eqalign{A&={1\over 4} C,\cr
B&=-{C\over 4},\cr
\phi&={\rm constant},\cr}}
with $C_o = 0$.  $C$ is given by
\eqn\fourc{e^{-C} = 1 + {k_4\over y^4}}
and
\eqn\kfour{k_4 = \kappa^2 T_4/2\Omega_5 ,}
The mass per unit three-volume of the threebrane is
\eqn\massfour{{\cal M}_4 = T_4 ,}
Once again, we find that this solution preserves half the spacetime
supersymmetries \dufltb\ and this establishes the existence of the self-dual
Type IIB superthreebrane.

In summary, for $D = 10$ we have found elementary and solitonic
string/fivebrane
solutions for $N = 1$, Type IIA and Type IIB; particle/sixbrane solutions,
membrane/fourbrane solutions for Type IIA
only and self-dual threebrane solutions for Type IIB only, all of which are
supersymmetric.

\newsec{\bf  $D = 11$}

We now turn our attention to $N = 1$, $D = 11$ supergravity.
Before doing so, however,
it is convenient to make the replacement \solution\ in \sigmact\ so that
\eqn\sigmactc{\eqalign{I_D (d) =&{1\over 2\kappa^2} \int d^Dx \sqrt{-g}
e^{-(D-2) \alpha^2C/4d}\bigg[R\cr
&-{\alpha^2\over 8} \bigg(1 - {\alpha^2 (D-1) (D-2)\over 2d^2}\bigg)
(\partial C)^2 -
{1\over 2 \cdot (d+1)!} F_{d+1}^2\bigg],\cr}}
where we have set $C_o = 0$ for simplicity.  If we now focus on the case
($D = 11$, $d= 3$) we find from \alphaexpre\ that
\eqn\elevenalpha{\alpha (3) = 0,}
and hence
\eqn\elevenact{I_{11} (3) = {1\over 2\kappa^2} \int d^{11}x \sqrt{-g}
\bigg[R - {1\over 2.4!}F_4^2\bigg].}
This is to be compared with the bosonic sector of $D = 11$ supergravity
\eqn\realeleven{I (D = 11 SUGRA) = {1\over 2\kappa^2} \int d^{11}x \sqrt{-g}
\bigg[R - {1\over 2.4!}F_4^2\bigg] - \int F_4 \wedge F_4 \wedge A_3.}
As discussed in section 7, there is no dualized form of this action since $A_3$
enters
explicitly.  We can however find elementary membrane solution.
Once again, this is just a special case of our general
solutions (3.15--18).
 For $d= 3$, $\tilde{d} = 6$, $\alpha (3) = 0$ we find explicitly
\eqn\elevensol{\eqalign{A&={1\over 3} C,\cr
B&=-{1\over 6} C,\cr}}
$C$ is given by
\eqn\elevenc{e^{-C} = 1 + {k_3\over y^6},}
and
\eqn\elevenk{k_3 = \kappa^2 T_3/3 \Omega_7.}
The mass per unit area of the membrane is
\eqn\elevenmass{{\cal M}_3 = T_3.}
This is Duff-Stelle \dufs\ solution which breaks half the supersymmetries and
 corresponds to the
eleven-dimensional
supermembrane of Bergshoeff, Sezgin and Townsend \berst.

\newsec{\bf  Double dimensional reduction and $D < 10$ supersymmetry.}

Simple dimensional reduction allows us to derive the actions
$I_D (d)$~and~$S_d$
 for a
($d - 1$)-brane moving in a $D$-dimensional spacetime from the actions
$I_{D + 1}(d)$~and~$S_d$ corresponding to a ($d - 1$)-brane in a
$(D + 1)$-dimensional spacetime.  This
corresponds to
\eqn\simpred{\eqalign{D + 1&\rightarrow D,\cr
d&\rightarrow d,\cr
\tilde{d} + 1&\rightarrow \tilde{d},\cr}}
and takes us vertically on the brane-scan.  Double dimensional reduction
\dufhis, on the other
hand, allows us to derive the actions $I_D (d)$~and~$S_d$ for a
($d - 1$)-brane moving
in $D$-dimensional spacetime from the actions
$I_{D + 1} (d + 1)$~and~$S_{d + 1}$.  This
corresponds to
\eqn\doubred{\eqalign{D + 1&\rightarrow D,\cr
d + 1&\rightarrow d,\cr
\tilde{d}&\rightarrow \tilde{d},\cr}}
and takes us diagonally on the brane-scan.  The first example of this was to
rederive the
Type IIA superstring in $D = 10$ from the supermembrane in $D = 11$ \dufhis.
 This
process
thus allows us, for example, to rederive the Dabholkar et al superstring
\dabghr\ solution in
$D = 10$ from the Duff-Stelle supermembrane \dufs\ solution in $D = 11$.

To see how it works in general, let us denote all ($D + 1, d + 1$)-dimensional
 quantities
by a hat and all ($D, d$) dimensional quantities without.  Then with
\eqn\split{\eqalign{\hat{X}^{\hat{M}}&= (X^M, X^d), \qquad M = 0, 1, \ldots,
 (d - 1),  (d +1), \ldots, D - 1\cr
\hat{\xi}^{\hat{\mu}}&= (\xi^i, \xi^d),\cr}}
double dimensional reduction consists in setting
\eqn\set{\xi^d = X^d,}
taking $X^d$ to be the coordinate on a circle of radius $R$, and discarding
all but the
zero modes.  In practice, this means taking the background fields $\hat{\phi}$,
$\hat{g}_{\hat{M}\hat{N}}$~and~$\hat{A}_{\hat{M}\hat{N} \ldots \hat{M}_d}$ to
be
independent of $X^d$.  To recover $S_d$, with only background fields $\phi$,
$g_{MN}$~and~$A_{M_1 M_2 \ldots M_{d-1}}$, a further truncation is necessary.
Specifically we write
\eqn\splitfield{\hat{g}_{\hat{M}\hat{N}} (\sigma-{\rm model})=
e^{-2 \beta\phi/d + 1}
\bigg(\matrix{g_{MN} (\sigma-{\rm model})&0\cr
 0&e^{2\beta\phi}\cr}\bigg),}
where $\beta$ is a, for the moment, arbitrary constant and
\eqn\reduten{\hat{A}_{0 1 2 \ldots d + 1} = A_{0 1 2 \ldots d},}
with other components set to zero.  The condition \splitfield\ ensures from
\defgama\ that
\eqn\relgama{\sqrt{-\hat{\gamma}} = \sqrt{- \gamma},}
and hence, together with condition \reduten, we recover the correct
$\sigma$-model
action for $S_{d - 1}$ starting from $\hat{S}_d$ provided
\eqn\redutension{ 2\pi R \hat{T}_{d + 1} =  T_d.}
We fix $\beta$ and the relation between $\hat \phi$ and $\phi$ by requiring
that we obtain the correct background field action
 $I_D (d)$
starting from $I_{D + 1} (d + 1)$. So from (6.17)
\eqn\reduricci{\eqalign{&e^{- (D - 1) \hat{\alpha} \hat{\phi}/2 (d + 1)}
\sqrt{-\hat{g}} \bigg[\hat{R} - {1\over 2}\bigg(1 - {{\hat\alpha}^2 ( D
(D - 1) \over 2 (d + 1)^2}\bigg)(\partial \hat {\phi})^2 \bigg] \cr
&= e^{-(D - 2) \alpha \phi/2d} \sqrt{-g} \bigg[R - {1 \over 2} \bigg(1 -
{\alpha^2 (D - 1)(D - 2)\over 2 d^2}\bigg) (\partial \phi)^2\bigg],\cr}}
which gives
\eqn\or{\eqalign{\hat{\phi} &= \delta \phi,\cr
 {(D - 1) \hat {\alpha}  \over 2 (d + 1)} \delta &= {(D - 2)\alpha
\over 2 d} - {\tilde d \beta \over d + 1},\cr
1 - {\alpha^2 (D - 1)(D - 2)\over 2 d^2}
&= \delta^2 \bigg(1 - {{\hat\alpha}^2  D (D - 1) \over 2 (d + 1)^2}
\bigg) \cr
&\quad - 4\beta {\tilde d + 1\over d + 1} {(D - 2) \alpha \over 2d}\cr
&\quad + 2 \beta^2 {D(D - 1) - 2(d + 1)(\tilde d + 1)\over (d + 1)^2},\cr}}
and hence
\eqn\betaexp{\beta = {2\over d \alpha },}
\eqn\gammaexp{\delta = {{\hat \alpha}\over \alpha},}
from solving eqs. \or.
We also require
\eqn\redukappa{\hat{\kappa}^2 = 2\pi R \kappa^2.}
Note that the Dirac quantization rule \tensiondirac\ involving
$\kappa^2$~and~$T$ follows from that
involving $\hat{\kappa}^2$~and~$\hat{T}$ on using \redutension\ and
\redukappa.  In canonical variables,
we have

\eqn\redumetric{\eqalign{\hat{g}_{MN} ({\rm canonical})&=
e^{-2\tilde{d}\phi/\alpha(d) (d + \tilde{d}) (d +1 + \tilde{d})}
g_{MN} ({\rm canonical}),\cr
\hat{g}_{dd} ({\rm canonical})&=
e^{{2\tilde d} \phi/(d + 1 +\tilde{d})\alpha (d)}.\cr}}

As an application of simultaneous dimensional reduction, we may derive the
Dabholkar et al
elementary string solution (8.3--5) in $D = 10$ from the Duff-Stelle
membrane solution in $D = 11$.
The $D = 10$ fields $g_{MN}$, $A_{MN}$~and~$\phi$ are given by
\eqn\duffstelle{\eqalign{\hat{g}_{MN}&= e^{-\phi/6} g_{MN} ({\rm canonical}),
\cr
\hat{g}_{22}&= e^{4\phi/3},\cr
\hat{A}_{012}&= A_{01}.\cr}}
[Curiously, the metric $\hat{g}_{MN}$ in \duffstelle\ bears the same relation
to $g_{MN}$ (canonical) as does the fivebrane $\sigma$-model metric in
\dusigmac\ since $\alpha (d = 2) = 1$~and~$\tilde{d} = 6$.  This phenomenon
happens in general whenever $\hat{\alpha} = 0$
i.e~~for ($d + 1 = 3, \tilde{d} = 6$), ($d + 1 = 4, \tilde{d} = 4$) and
($d + 1 = 6,\tilde{d} = 3$)].  Similarly starting from sixbrane in $D = 10$ we
 may proceed diagonally
down the brane-scan to a particle in $d = 4$.  It is not difficult to show that
the
solutions so obtained will continue to preserve exactly one half of the
supersymmetries.
Starting from the $d \leq 7$ solutions in $D = 10$ we can thus fill out the
triangle of
supersymmetric extended objects described in \duflbs.

\newsec{\bf   $D = 6$:  The self-dual string}

If we proceed by double dimensional reduction from the super-fivebrane in
$D = 10$ we arrive at a super $(\tilde{d} = 2)$ string in $D = 6$ for which
$\alpha (\tilde{d} = 2) = -\alpha (d = 2)$ i.e which is dual to the elementary
 super string and related by a
strong/weak coupling replacement $\phi \rightarrow - \phi$. Compare (6.23)
with (6.24).

However, there is another supersymmetric solitonic string in $D = 6$:  the
\underbar{self-dual}\break\hfil \underbar{superstring} which falls outside our
previous
discussions and requires a special treatment.  This is the $D = 6$ counterpart
of the
self-dual superthreebrane in $D = 10$.  Our starting point is the $N = 2$,
$D = 6$ self-dual
supergravity \refs{\rom,\sala} which, in common with the Type IIB superstring
in $D =10$,
 admits
covariant field equations, but no manifestly covariant field equations.  It
describes a
graviton $e_M\,^A$, two left-handed gravitini $\psi_{Ma}$ and one tensor field
 $B_{MN}$ with
self-dual field strength $G_{MNP}$.  The gravitini transformation rules are
(in our notation)
\eqn\gtinitransf{\delta \psi_M = \nabla_M \varepsilon - {1\over 8}
G_{MNP} \Gamma^{NP}\varepsilon.}
So if we make a two/four split as in section 3 with
\eqn\splitgama{\eqalign{\Gamma_A&= (\gamma_{\alpha} \otimes 1, \gamma^3
\otimes \Sigma_m), \qquad \Gamma^7 = \gamma^3 \otimes \Gamma^5,\cr
\gamma^3&=\gamma^0 \gamma^1, \qquad \Gamma^5 = \Sigma^2 \Sigma^3 \Sigma^4
\Sigma^5,\cr}}
the criterion for unbroken supersymmetry, $\delta \psi_M = 0$, reduces to
\eqn\criterion{\eqalign{\partial_{\mu} \varepsilon - {1\over 2} \gamma^3
\gamma_\mu \otimes \Sigma^n (\partial_n A
+ {1\over 2} e^{-2A} \partial_n e^C \gamma^3) \varepsilon&=0,\cr
\partial_m \varepsilon + {1\over 2} \partial_m B \varepsilon - {1\over 2}
(\delta^n\,_m +
\Sigma^n\,_m) (\partial_n B - {1\over 2} e^{-2A} \partial_m e^C \gamma^3)
\varepsilon&=0,\cr}}
and hence supersymmetry requires
\eqn\susygiving{C = 2A, \qquad B = - A, \qquad \varepsilon = e^{-B/2}
\varepsilon_o,}
where $\varepsilon_o$ obeys $\gamma^3 \varepsilon_o = - \varepsilon_o$, and
one half of the
supersymmetries is broken.

The bosonic equations of motion are
\eqn\einsteineq{R_{MN} - {1\over 2} g_{MN} R = {1\over 4}
G_M\,^{PQ} G_{NPQ}}
\eqn\selfdueq{G_{MNP} =- \tilde{G}_{MNP},}
and substituting \susygiving\ yields
\eqn\munucomp{e^{6A} \delta^{mn}\partial_m \partial_n e^{-2A} = 0,}
for the $\mu\nu$ components of the Einstein equation and
\eqn\mncomp{e^{2A} \delta^{mn} \partial_m \partial_n e^{-2A} = 0,}
for the $mn$ components.  So
\eqn\solution{e^{-2A} = 1 + {k_2\over y^2}.}
All the properties of the dyonic self-dual threebrane \dufltb\ apply,
mutatis mutandis, to the
dyonic self-dual string, including Dirac quantization rules and the saturation
 of the
Bogolmol'nyi bound.

The effective bosonic equations of motion of this string are
\eqn\effecact{\eqalign{&\partial_i (\sqrt{-\gamma} \gamma^{ij}
\partial_j X^N g_{MN}) - {1\over 2}
\sqrt{-\gamma} \gamma^{ij} \partial_i X^N \partial_j X^P \partial_M g_{NP}\cr
&={1\over 2} G_{MNP} \partial_i X^N \partial_j X^P \varepsilon^{ij},\cr}}
but, since $G_{MNP} = - \tilde{G}_{MNP}$, there is no manifestly covariant
world
sheet
action.  It would be interesting to include the fermionic degrees of freedom
and
 construct
the spacetime supersymmetric, $\kappa$-symmetric,  Green-Schwarz string
equations, but this
has not yet been done.

\newsec{\bf   $D = 4$: Electric-Magnetic Duality}

Alternatively, we may proceed vertically down the brane-scan as far as
$\tilde{d} = 1$.
Thus starting with a particle in $D = 10$ we arrive at a particle in $D = 4$.
However,
this solution will have an $\alpha$ parameter $\alpha (d = 1) = + \sqrt{3}$
opposite in
sign to
the $\alpha(\tilde d = 1) = - \sqrt{3}$ solution obtained by double
dimensional reduction from
 the
sixbrane in $D = 10$.  The two Lagrangians are given by
\eqn\particleact{I_4 (1) = {1\over 2\kappa^2} \int d^4 x \sqrt{-g}
\bigg(R - {1\over 2} (\partial\phi)^2
- {1\over 4} e^{-\sqrt{3} \phi} F_{\mu\nu} F^{\mu\nu}\bigg),}
\eqn\dualpact{\tilde{I}_4 (1) = {1\over 2\kappa^2} \int d^4 x \sqrt{-g}
\bigg(R - {1\over 2}(\partial\phi)^2 - {1\over 4} e^{+ \sqrt{3} \phi}
\tilde{F}_{\mu\nu}\tilde{F}^{\mu\nu}\bigg),}
where
\eqn\dualtwo{\tilde{F}_{\mu\nu} = e^{- \sqrt{3} \phi}\,^{\ast}F_{\mu\nu}.}
If, for simplicity, we set $\phi_o = 0$, then the action $I_4 (1)$ admits the
elementary solution
\eqn\elemenpar{\eqalign{ds^2&=- \bigg(1 + {k_1\over y}\bigg)^{-1/2} dt^2 +
\bigg(1 + {k_1\over y}\bigg)^{1/2} (dy^2 + y^2 d\theta^2 + y^2
\sin^2 \theta d\phi^2),\cr
e^{2\phi}&= \bigg(1 + {k_1\over y}\bigg)^{- \sqrt{3}},\cr
B_o&= - \bigg(1 + {k_1\over y}\bigg)^{-1},\cr}}
where
\eqn\kappaone{k_1 = {\kappa^2 T_1\over 2\pi},}
corresponding to an electric monopole with mass
\eqn\parmass{m = T_1,}
and electric charge
\eqn\parcharge{e = \sqrt{2} \kappa m.}
$I_4 (1)$ also admits the solitonic solution
\eqn\solipar{\eqalign{ds^2&= - \bigg(1 + {k_{\tilde{1}}\over y}\bigg)^{-1/2}
dt^2 + \bigg(1 + {k_{\tilde 1}\over y}\bigg)^{1/2} (dy^2 + y^2 d\theta^2 +
y^2 \sin^2 \theta d\phi^2),\cr
e^{2\phi}&= \bigg(1 + {k_{\tilde{1}}\over y}\bigg)^{\sqrt{3}},\cr
{1\over \sqrt{2} \kappa} F_{\theta\phi}&= {g \over 4\pi},\cr}}
where
\eqn\tildekappaone{k_{\tilde{1}} = {\kappa^2 \tilde{T}_1\over 2\pi} =
{n\over 2 T_1},}
corresponding to a magnetic monopole with mass $\tilde{T_1}$  and magnetic
 charge
obeying
\eqn\parchargequan{e g = 2\pi n.}
For the dual action $\tilde{I}_4 (1)$, the electric and magnetic solutions are
 interchanged.

These solutions are precisely the extreme mass = charge limits of the
black-hole
 solutions
of $D = 4,~N = 8$ supergravity discussed by Gibbons and Perry \gibp.
 See also Han et al
\han\ who obtained the same solution from $D = 11$ supergravity.  Gibbons and
Perry pointed
out that, considered as solutions of $N = 8$ supergravity in $D = 5$, the
monopole solitons
fit into the same supermultiplets as the elementary electric monopoles, and
went
 on to
speculate that there exists a dual theory for which the roles of elementary and
solitonic
particles are interchanged.  In the light of the results of the present paper,
 we may
re-interpret this electric-magnetic duality conjecture in $D = 4$ as a
particle/sixbrane
duality conjecture in $D = 10$.  (Note that the values of $\alpha$ considered
here i.e $\pm\sqrt{3}$ differ from the $\alpha = \pm 1$ considered by
Shapere et al \shatw\ and
Kallosh et al \kallopv\ in the same context of electric-magnetic duality and
to which, therefore, the above remarks do not apply.)

\newsec{\bf  $D = 4$  axionic instanton}

Another special case of interest corresponds to $\tilde d = 0$, $D = 4$,
 $\alpha (2) = 2$ (solitonic solution).  Since there is now no time
coordinate, this corresponds to a Euclidean instanton.
{}From \solisolution, we have
\eqn\instanton{\eqalign{ds^2&=\delta_{mn} dy^m dy^n, \qquad m = 1, 2, 3, 4
\cr
 e^\phi&=1 + {k_o\over y^2}.\cr
{1\over \sqrt{2} \kappa}~F_3&= g_0 \varepsilon_3/\Omega_3\cr}}
The metric is flat, and the energy-momentum tensor vanishes.  This
solution is  just the axionic instanton first discovered by Soo-Jong Rey \rey.

\newsec{\bf  Black branes}

In the case $D = 10$, Horowitz and Strominger discovered two parameter
solitonic solutions
of the theory \lgeact\ in the cases $1 \leq d \leq 7$ which displayed event
horizons:  the
``black p-branes'' \hors.  The two parameters are the mass per unit volume
${\cal M}_{\tilde{d}}$ and the charge per unit volume $g_{\tilde{d}}$,
which satisfy the
Bogolmol'nyi bound $\kappa {\cal M}_{\tilde{d}} \geq g_{\tilde{d}}/\sqrt{2}$.
In this
section, we generalize their results to arbitrary $D$.  We also show that in
the
 cases $1\leq d \leq 7$~and~$1 \leq \tilde{d} \leq 7$,
the {\it extreme} black p-branes i.e those
obeying the mass $=$ charge limit
$\kappa {\cal M}_{\tilde{d}} = g_{\tilde{d}}/\sqrt{2}$,
coincide with the Type II super p-branes \duflbs.
\vfill\eject
Using the canonical metric, the $(\tilde{d} - 1)$ brane black soliton solution
may be
written for all $\tilde{d} \geq 1$ as
\eqn\blacksolution{\eqalign{ds^2=& - \bigg[1 - \bigg({r_+\over r}\bigg)^d\bigg]
 \bigg[1 - \bigg({r_-\over r}\bigg)^d\bigg]^{-\tilde{d}/(d + \tilde{d})}
dt^2\cr
&+ \bigg[1 - \bigg({r_+\over r}\bigg)^d \bigg]^{-1} \bigg[1 -
\bigg({r_-\over r}\bigg)^d \bigg]^{{\alpha^2\over 2d} - 1} dr^2\cr
&+ r^2 \bigg[1 - \bigg({r_-\over r}\bigg)^d\bigg]^{{\alpha^2\over 2d}}
 d \Omega^2_{d + 1}\cr
&+ \bigg[1 - \bigg({r_-\over r}\bigg)^d \bigg]^{{d\over d + \tilde{d}}}
dx^i dx_i, \qquad i = 1 \ldots \tilde{d} - 1,\cr
e^{-2\phi}&= \bigg[1 - \bigg({r_-\over r}\bigg)^d\bigg]^{\alpha (d)},\cr
{1\over \sqrt{2} \kappa} F_{d + 1}&=g_{\tilde{d}} \varepsilon_{d + 1}/
\Omega_{d + 1},\cr}}
where the magnetic charge $g_{\tilde{d}}$ and the mass per unit
$(\tilde{d} - 1)$-volume
${\cal M}_{\tilde{d}}$ are related to $r_{\pm}$ by \lu\
\eqn\bmcharge{g_{\tilde{d}} = {\Omega_{d + 1}\over \sqrt{2} \kappa}
d (r_+ r_-)^{d/2},}
\eqn\bmass{{\cal M}_{\tilde{d}} = {\Omega_{d + 1}\over 2 \kappa^2}
[(d + 1) r_+^d - r_-^d].}
We note that there are consistent with the Bogolmol'nyi bound \magbogom\ with
$\phi_o = 0$.  The
solutions poses an $R \times SO (d + 2) \times E (\tilde{d} - 1)$ symmetry
where
 $E (n)$
denotes the $n$-dimensional Euclidean group.  The solutions exhibit an event
horizon at $r= r_+$ and an inner horizon at $r = r_-$.  In the special case
$D = 11$, $\tilde{d} = 3, 6$
they reduce to the black membrane and black fivebrane of Guven \guv.
 In the special case $D = 10$ i.e $\tilde{d} = 8 - d$, they reduce to the
Horowitz-Strominger black $p$-brane
solutions \hors.  In the special case $D = 4$, $\tilde{d} = 1$ they reduce to
the
dilaton
black hole solution of Gibbons and Perry \gibp.  Two special cases of interest
 are the zero
charge solutions $(r_- = 0)$ and the extreme mass $=$ charge solutions
$(r_+ = r_-)$.  In
the first case the dilaton and antisymmetric tensor are trivial and the metric
 reduces to
\eqn\trivialmetric{\eqalign{ds^2&=-V dt^2 + V^{-1} dr^2 + r^2
d\Omega_{d + 1}\,^2 + dx^i dx_i,\cr
V&=1 - \bigg({r_+\over r}\bigg)^d .\cr}}
Gregory and Laflamme have argued that (in the $D = 10$ case) these solutions
are classically unstable \grel.  In the second case, remarkably enough,
at the external limit
$r_+ = r_-$ the metric component $g_{oo}$ becomes equal to the one multiplying
$dx^i dx^i$
and the symmetry is enlarged to $SO (d + 2) \times P (\tilde{d})$:
\eqn\extresolu{\eqalign{ds^2=&\bigg[1 - \bigg({r_-\over r}\bigg)^d
\bigg]^{d/(d + \tilde{d})} dx^{\mu}
 dx_{\mu}\cr
&+\bigg[1 - \bigg({r_-\over r}\bigg)^d\bigg]^{\alpha^2\over 2d}
\bigg\{\bigg[1 - \bigg({r_-\over r}\bigg)^d\bigg]^{-2} dr^2 + r^2
d\Omega^2_{d + 1}\bigg\},\cr
e^{-2\phi}=&\quad \bigg[1 - \bigg({r_-\over r}\bigg)^d\bigg]^{\alpha(d)},\cr
{1\over \sqrt{2} \kappa} F_{d + 1}=&\quad g_{\tilde{d}}
\varepsilon_{d + 1}/\Omega_{d +1}.\cr}}
It is convenient to introduce the change of variables $y^d = r^d - r_-^d$,
then \extresolu\ becomes
\eqn\extreme{\eqalign{ds^2 &=\bigg[1 + \bigg({r_-\over y}\bigg)^d
\bigg]^{-d/(d + \tilde{d})} dx^{\mu} dx_{\mu}
+ \bigg[1 + \bigg({r_-\over y}\bigg)^d\bigg]^{\tilde{d}/(d + \tilde{d})}
(dy^2 + y^2 d\Omega^2_{d +1}),\cr
e^{-2\phi}&=\bigg[1 + \bigg({r_-\over y}\bigg)^d\bigg]^{-\alpha(d)},\cr
{1\over \sqrt{2} \kappa} F_{d + 1}&=g_{\tilde{d}} \varepsilon_{d + 1}
/\Omega_{d + 1}.\cr}}
But in the case $1 \leq d \leq 7$, $1 \leq \tilde{d} \leq 7$, these are
precisely the super $p$-branes, so $r_+ = r_-$ also corresponds to the
appearance of supersymmetry.

It is also possible to find {\it elementary} black $(d - 1)$-branes with
parameters ${\cal M}_d$~and~$e_d$ obeying the bound
$\kappa {\cal M}_d \geq \sqrt{2} e_d$, by including a
source term on the right hand side of the equations.  In this case however,
it would be
necessary to relax the equality of the kinetic and WZW term coefficients in
\gepbact\ to allow
for mass $\not=$ charge.  (This equality is forced on us in the supersymmetric
 case, by
virtue of $\kappa$-symmetry \berst).
\vfill\eject

\newsec {\bf   Black and super p-branes:  Singular or non-singular?}

In this section, we would like to classify the singular nature of black and
super $p$-branes
discussed in previous sections thus generalizing the results of \dufkl.
The black ($d -1$)-brane solutions can be obtained from the black
($\tilde{d} - 1$)-brane solutions simply
by sending $d \leftrightarrow \tilde{d}$ in previous section.  The Ricci scalar
of black
($d - 1$)-brane calculated in terms of ($n - 1$)-brane variables is
\eqn\ricciscalar{\eqalign{R&= \bigg[1 - \bigg({r_-\over r}\bigg)^{\tilde{d}}
\bigg]^{-\big({\alpha^2 (d)\over
2\tilde{d}} + {\alpha (d) \alpha (n)\over 2n}\big)} \times {Q^2\over r^{2
(\tilde {d} + 1)}}\cr
&\bigg\{{1\over 8}\bigg[2 (\tilde{d} + 2) (1 - \tilde{d}) - {(D - 1) (D - 2)
\over 2}
\bigg({\alpha^2 (d) \alpha^2 (n)\over n^2} - {4\tilde{d}^2\over (D - 2)^2}
\bigg)
\bigg]\cr
&\bigg({r_-\over r_+}\bigg)^{\tilde{d}} {1 - ({r_+\over r})^{\tilde{d}}\over
1 - ({r_-\over r})^{\tilde{d}}}\cr
&+{1\over 2} \bigg[2\tilde{d} + 1 - (D - 1)
\bigg({\alpha (d) \alpha (n)\over n} +
{2\tilde{d}\over D - 2}\bigg)\bigg]\bigg\}.\cr}}

We note that for either $n = d$ or $n = \tilde d$, i.e. written in its
own variables or its dual ones,  the R in
\ricciscalar\ always
 blows up as
r goes to $r_-$.  The reason is that
$S \equiv -{\alpha^2 (d)\over 2\tilde{d}} - {\alpha (n) \alpha (d)\over 2n}$
is now  less than one.  For $n = d$,
$S = - \half ({1\over d} + {1\over \tilde d}) \alpha^2 (d) < 0$, and for
$n = \tilde d$, S = 0 since $\alpha (\tilde d) = - \alpha(d)$. In conclusion,
physically interesting
black p-branes always display singularities at $r = r_-$.

The situation for super $p$-branes is quite different.
The calculations proceed along the
same lines as \dufkl.  Here we omit the details and simply state the results.
  We can always
choose suitable variables to get rid of the singularities (for example, the
dual variables)
or else there is no singularity at all (as for example in the self-dual
threebrane).  We can
also calculate
the proper time for a ($n - 1$)-brane falling into a ($d - 1$)-brane.  We find
that only for
strings and their dual objects, the corresponding proper time is infinite,
which
agrees with what we have discussed in \dufkl.  For any other object and
its dual, the corresponding proper
time is always finite although there is no curvature singularity when written
in terms of the
dual variables, which contradicts our naive expectations.  For the self-dual
cases, the
self-dual threebrane \dufltb\ is free of singularity and the corresponding
proper time is
finite, and similarly for the self-dual string.  Any extended object, except
 for self-dual
string
and threebrane, has a curvature singularity when written in terms of its own
variables
 and the
corresponding proper time is finite.  From the above, it is easy to see that
only strings
(except for the self-dual one) satisfy our naive expectations.

\appendix A {\bf Comparison with Brans-Dicke Theory}

The action for Brans-Dicke gravity (generalized from 4 to D dimensions) may
be written in
terms of a scalar field $\eta$ and some metric $g_{MN}$ (BD)
\eqn\appenone{I ({\rm Brans-Dicke}) = {1\over 2\kappa^2} \int d^D x \sqrt{-g}
\bigg[\eta R -{\omega\over \eta} (\partial\eta)^2\bigg] +
\int d^D x {\cal L} ({\rm matter},g),}
where $\omega$ is a free parameter and where, by construction,
${\cal L}$ (matter, $g$) is
independent of $\eta$.  In comparing this to our general action $I_D (d)$ we
have to decide
what is meant by ${\cal L}$ (matter, $g$).  Let us first suppose that this
refers not to
the antisymmetric tensor action of \lgeact\ but to the ($d - 1$)-brane action
 $S_d$ of \gepbact.
Then we must make the identification
\eqn\appentwo{g_{MN} (BD) = g_{MN} (d),}
where $g_{MN} (d)$ is the ($d - 1$)-brane $\sigma$-model metric of \sigmacano.
 Comparison with
\sigmact\ then yields the identifications
\eqn\appenthree{\eta = e^{-(D - 2) \alpha (d) \phi/2d},}
\eqn\appenfour{\omega = {2d^2\over (D - 2)^2 \alpha^2 (d)} -
{D - 1\over D - 2} = - {(D - 1) (d - 2) -d^2\over (D - 2) (d - 2) - d^2},}
where we have used \alphaexpre.  It is interesting to note, for example, that
in $D = 10$
strings ($d = 2$) correspond to $\omega = - 1$, fivebranes ($d = 6$) to
$\omega = 0$ and
threebranes ($d = 4$) to $\omega = \infty$.  However, if we now include in
${\cal L}$
(matter, $g$) the action for a ($d' - 1$) antisymmetric tensor written in ($d
-
1$)-brane
variables, namely

$${1\over 2\kappa^2} \int d^D x \sqrt{-g} e^{-(D - 2) \alpha (d) \phi/2d}
e^{[d'\alpha (d)
- d\alpha (d')]\phi/d} {1\over 2 (d' + 1)!} F^2_{d' + 1}\eqno(A.5)$$

\noindent
then our theory will no longer be of the Brans-Dicke form unless

$$d' \alpha (d) - d\alpha (d') = \bigg({D - 2\over 2}\bigg) \alpha
(d)\eqno(A.6)$$

\noindent
If $d = d'$ this will not be satisfied unless $(D - 2) \alpha (d)$ vanishes in
which case
$\omega = \infty$.  If, on the other hand, we omit the action $S_d$ and take
${\cal L}$
(matter, $g$) to indicate the antisymmetric tensor action alone, then we may
re-interpret
(A.5) as the tree-level action for a ($d' - 1$)-brane written in $\sigma$-model
variables.
Then (A.6) is no longer a restriction on the dimension of the extended object
but only on
the variables we choose to write the action.

\bigbreak\bigbreak\bigbreak\centerline{{\bf Acknowledgments}}\nobreak We would
like to thank J. Dixon, G. Horowitz, R. Khuri, H. La, R. Minasian, J.
Rahmfeld,
E. Sezgin and A. Strominger for helpful discussions.

\listrefs

\end